\documentclass[10pt,a4paper]{article}
\usepackage[english]{babel}

\usepackage{amsmath}
\usepackage{amssymb}
\usepackage{graphicx,epsfig}
\usepackage[mathscr]{eucal}

\usepackage{color}
\usepackage[numbers,sort&compress]{natbib}
\usepackage{multirow}
\usepackage{tabularray}
\usepackage{tabularx}
\UseTblrLibrary{booktabs}
\usepackage{float}

\usepackage{slashed}
\usepackage{braket}

\usepackage{scalefnt,ulem,pstricks}

\usepackage{xspace}
\usepackage{setspace}
\usepackage{xstring}
\usepackage{mathrsfs}
\usepackage{amsbsy}

\def\ltap{\raisebox{-.4ex}{\rlap{$\,\sim\,$}} \raisebox{.4ex}{$\,<\,$}}
\def\gtap{\raisebox{-.4ex}{\rlap{$\,\sim\,$}} \raisebox{.4ex}{$\,>\,$}}

\newcommand{\M}{\mathcal{M}}

\newcommand{\J}{\mathcal{J}}
\newcommand{\SSoft}{\mathbf{S}}
\newcommand{\Soft}{\pmb{\mathscr{S}}}
\newcommand{\Hard}{\mathcal{H}}

\newcommand{\ZNeubert}{\mathbf{Z}}

\newcommand{\finSCET}{\mathrm{fin}}

\newcommand{\alphasNf}{\alpha_{\rm S}^{\!(n_f)}}
\newcommand{\alphasNl}{\alpha_{\rm S}^{\!(n_l)}}
\newcommand{\as}{\alpha_{\rm S}}
\newcommand{\MSbar}{\overline{\mathrm{MS} }}

\newcommand{\SA}{\mathrm{SA}}
\newcommand{\MA}{\mathrm{MA}}

\def\ttbH{\ensuremath{t {\bar t}H}\xspace}

\def\ttH{\ttbH}
\def\ttW{\ensuremath{t \bar t W}\xspace}

\def\muF{{\mu_{\mathrm{F}}}}
\def\muR{{\mu_{\mathrm{R}}}}
\def\muIR{{\mu_{\mathrm{IR}}}}
\def\rcut{r_{\mathrm{cut}}}

\newcommand{\HQQAmp}{{\scshape HQQAmp}\xspace}
\newcommand{\Pentagon}{{\scshape PentagonFunctions++}\xspace}
\newcommand\Matrix{{\sc Matrix}\xspace}

\newcommand\OpenLoops{{\sc OpenLoops}\xspace}
\newcommand\Recola{{\sc Recola}\xspace}

\newcommand\Whizard{{\sc Whizard}\xspace}

\usepackage{scalefnt,pstricks}
\usepackage{cancel}

\usepackage{array}
\newcolumntype{L}[1]{>{\raggedright\let\newline\\\arraybackslash\hspace{0pt}}m{#1}}
\newcolumntype{C}[1]{>{\centering\let\newline\\\arraybackslash\hspace{0pt}}m{#1}}
\newcolumntype{R}[1]{>{\raggedleft\let\newline\\\arraybackslash\hspace{0pt}}m{#1}}

\usepackage[font=normal, labelfont=bf]{caption}

\usepackage{lipsum}
\usepackage[colorlinks=true,allcolors={blue!70!black}]{hyperref}

\begin{document}
\begin{titlepage}
\begin{flushright}
ZU-TH-57/24 \\
PSI-PR-24-22 \\
TUM-HEP-1538/24
\end{flushright}

\renewcommand*{\thefootnote}{\fnsymbol{footnote}}
\vspace*{0.5cm}

\begin{center}
  {\Large \bf Precise predictions for $\boldsymbol{\ttbH}$ production at the LHC:\\[0.2cm] inclusive cross section and differential distributions}
\end{center}

\par \vspace{2mm}
\begin{center}
    {\bf Simone Devoto${}^{(a)}$}, {\bf Massimiliano Grazzini${}^{(b)}$},\\[0.25cm]
    {\bf Stefan Kallweit${}^{(b)}$}, {\bf Javier Mazzitelli${}^{(c)}$} and {\bf Chiara Savoini${}^{(d)}$}

  \vspace{5mm}

${}^{(a)}$Department of Physics and Astronomy, Ghent University, 9000 Ghent, Belgium\\[0.25cm]

${}^{(b)}$Physik Institut, Universit\"at Z\"urich, 8057 Z\"urich, Switzerland\\[0.25cm]

${}^{(c)}$PSI Center for Neutron and Muon Sciences, 5232 Villigen PSI, Switzerland\\[0.25cm]

${}^{(d)}$Technical University of Munich, TUM School of Natural Sciences, Physics Department, James-Franck-Stra{\ss}e 1, 85748 Garching, Germany

  \vspace{5mm}

\end{center}

\par \vspace{2mm}
\begin{center} {\large \bf Abstract}

\end{center}
\begin{quote}
\pretolerance 10000

We present the first fully differential next-to-next-to-leading order (NNLO) QCD calculation for the production of a top--antitop quark pair in association with a Higgs boson (\ttH) at hadron colliders. 
The computation is exact, except for the finite part of the two-loop virtual contribution, which we estimate using two different methods that yield consistent results within their respective uncertainties.
The first method relies on a soft-Higgs factorisation formula that we develop up to the three-loop order. The second is based on a high-energy expansion in the small top-mass limit.
Combining the newly computed corrections with the complete set of next-to-leading order (NLO) QCD+EW results provides the most advanced perturbative prediction currently available at the LHC for both inclusive and differential \ttH cross sections. The uncertainties due to the missing exact two-loop contribution are conservatively estimated to be at the percent level, both for the total cross section and for most of the differential distributions, and therefore subleading compared to the residual perturbative uncertainties.

\end{quote}
\vskip 2cm
\begin{center}
{\it \large This paper is dedicated to the memory of Stefano Catani,\\[0.3cm] outstanding scientist and sincere friend}
\end{center}

\vspace*{\fill}
\end{titlepage}

\tableofcontents

\section{Introduction}
\label{sec:intro}
Twelve years after its discovery~\cite {ATLAS:2012yve,CMS:2012qbp}, the LHC data have significantly refined the Higgs boson picture,
by combining information from different production and decay modes. 
A coherent framework of Higgs boson interactions with heavy fermions ($t, b$, and $\tau$) and bosons ($W, Z$) has emerged, entirely consistent with the Standard Model (SM) hypothesis~\cite{ATLAS:2022vkf,CMS:2022dwd}.

Despite the success of the SM, many questions still need to be answered, with the Higgs sector offering a potential gateway to New Physics. 
One of the key objectives of the High-Luminosity (HL) LHC upgrade is to achieve even more precise measurements of the Higgs boson properties.
In this context, the Higgs boson coupling to the top quark plays a unique role, as it is expected to be of the order of unity and sensitive to effects beyond the SM.

The top-quark Yukawa coupling can be directly measured by studying the process in which a Higgs boson is produced in association with a top--antitop quark pair (\ttH).
Although this process accounts for only about one percent of the events compared to the dominant gluon fusion production mode, it has been successfully studied at the LHC, with the ATLAS and CMS collaborations reporting the first observation in 2018~\cite{Aaboud:2018urx,Sirunyan:2018hoz}.
The study of differential distributions is also envisaged, and a measurement of the transverse momentum of the Higgs boson was already reported in Refs.~\cite{ATLAS:2021qou,CMS:2024fdo,ATLAS:2024gth}.
Any deviation from the SM prediction would be a signal of New Physics.
Currently, ATLAS and CMS measure the signal strength in this channel with an accuracy of $\mathcal{O}(20\%)$. 
However, uncertainties are expected to decrease to $\mathcal{O}(2\%)$~\cite{Cepeda:2019klc} at the end of the HL phase, which will require an excellent control of the background modelling.
Precise theory predictions for this production mode, both at the inclusive and differential levels, will be crucial to match the anticipated improvements.

The first theoretical studies of \ttH production in the SM were carried out a long time ago in Refs.~\cite{Ng:1983jm,Kunszt:1984ri} at leading order (LO) in QCD perturbation theory, and in Refs.~\cite{Beenakker:2001rj,Beenakker:2002nc,Reina:2001sf,Reina:2001bc,Dawson:2002tg,Dawson:2003zu} at next-to-leading order (NLO).
These calculations have been matched with parton showers in Refs.~\cite{Frederix:2011zi,Garzelli:2011vp,Hartanto:2015uka}.
The NLO electroweak (EW) corrections have first been calculated in Ref.~\cite{Frixione:2014qaa} and then combined with NLO QCD contributions, within the narrow-width-approximation for the top-quark decays, in Refs.~\cite{Yu:2014cka,Frixione:2015zaa}.

The resummation of soft-gluon contributions close to the partonic threshold was considered in Refs.~\cite{Kulesza:2015vda,Broggio:2015lya,Broggio:2016lfj,Kulesza:2017ukk}, while soft and Coulomb corrections have been resummed in Ref.~\cite{Ju:2019lwp}. 
The resummed results up to next-to-next-to-leading logarithmic (NNLL) accuracy, matched to the fixed-order NLO results, have been further improved by also including EW corrections ~\cite{Broggio:2019ewu,Kulesza:2020nfh}.
The NLO QCD corrections to off-shell top quarks in $\ttH$, with leptonic $W$ decays, have been considered in Refs.~\cite{Denner:2015yca, Stremmer:2021bnk}, and full off-shell effects with \mbox{$H \to b \bar{b}$} decay in Refs.~\cite{Denner:2020orv,Bevilacqua:2022twl}.
A combination of the NLO QCD and EW effects has been presented in Ref.~\cite{Denner:2016wet}.

Given the projection that statistical uncertainties will shrink to the few-percent level, the $\ttH$ analyses will become dominated by systematics. Currently, the largest source of these uncertainties arises from theory modelling, with the $\ttH$ signal being characterised by a theoretical uncertainty of $\mathcal{O}(10\%)$~\cite{LHCHiggsCrossSectionWorkingGroup:2016ypw}.
Therefore, including next-to-next-to-leading order (NNLO) QCD corrections is essential to match the experimental precision expected by the end of the HL-LHC phase.

An important step forward was made in Ref.~\cite{Catani:2021cbl}, where NNLO QCD corrections for the flavour off-diagonal partonic channels were presented.
A complete NNLO calculation, including the diagonal partonic channels, requires
the evaluation of several infrared (IR) divergent components: tree-level contributions with two additional
unresolved partons, one-loop contributions with one unresolved parton, and purely virtual corrections.
While the required tree-level and one-loop scattering amplitudes can nowadays be evaluated using automated tools, 
the two-loop amplitude for \ttbH production remains unknown and represents a formidable task, as its calculation pushes the limits of current computational techniques for \mbox{$2 \to 3$} scattering processes. 
Indeed, the complexity of multi-loop scattering amplitudes rapidly increases with the number of loops and physical scales such as kinematic invariants and particle masses. 

In the past few years, impressive advances in understanding the analytic structure of scattering amplitudes have led to the computation of two-loop five-point amplitudes for many phenomenologically relevant processes. 
Exact results are available for several \mbox{$2 \to 3$} massless processes~\cite{Badger:2019djh,Agarwal:2021vdh,Badger:2021imn,Badger:2023mgf,Abreu:2023bdp,Agarwal:2023suw,DeLaurentis:2023nss,DeLaurentis:2023izi} while results in leading-colour (large $N_c$) approximation have been presented for few processes involving one external mass~\cite{Badger:2021nhg,Badger:2021ega,Abreu:2021asb,Badger:2022ncb,Badger:2024sqv}.
To date, the computation of a two-loop five-point amplitude with two or more massive external legs has yet to be completed.

The analytic complexity escalates dramatically when massive particles are involved in internal loops. 
This occurs because integrals associated with nontrivial algebraic curves and surfaces, such as elliptic curves, can arise~\cite{FebresCordero:2023gjh,Badger:2024fgb}.
In the context of \ttH production, this was pointed out in Ref.~\cite{FebresCordero:2023gjh}, where analytic results for the master integrals (MIs) contributing to the leading-colour two-loop amplitudes, proportional to the number of light flavours, were presented.
The presence of elliptic sectors represents an additional complication: solutions of the system of differential equations in terms of iterated integrals, with well-defined properties and suited for an efficient numerical evaluation, may not exist. 
One alternative could be a semi-analytic approach using generalised power series expansions~\cite{Moriello:2019yhu}. 
This method has been adopted in Ref.~\cite{Buccioni:2023okz}, where the \mbox{$gg \to t\bar{t}H$} one-loop amplitudes have been computed up to the second order in the dimensional regularisation parameter.
However, this strategy often results in prohibitive times for an on-the-fly numerical evaluation of the MIs and can lead to instabilities across the phase space.
Alternatively, a fully numerical approach has been employed in Ref.~\cite{Agarwal:2024jyq} to obtain the first numerical results for the quark-initiated \ttH two-loop amplitudes containing closed fermion loops.
Although the authors of Ref.~\cite{Agarwal:2024jyq} demonstrated that the computation achieves sufficient precision (\mbox{$\sim1\%$}) at individual phase space points, an on-the-fly evaluation remains impractical for realistic phenomenological applications. 
Consequently, the main challenge lies in constructing a densely sampled five-dimensional grid.

Despite substantial recent advancements, the computation of the two-loop \ttH amplitudes is not expected to be completed in the near future.
For practical phenomenological applications, a promising strategy consists of obtaining the missing two-loop amplitudes in an approximate form and assessing their impact on the NNLO cross section.
This approach was first applied in Ref.~\cite{Catani:2022mfv}, where the two-loop hard matrix elements have been computed in the soft Higgs boson limit.
The \textit{soft-Higgs approximation} provides considerable simplifications since the amplitude factorises in this limit. 
Its validity at the inclusive level is reinforced by a posteriori observation that the quantitative impact of the genuine two-loop corrections in the soft-Higgs limit is relatively small (\mbox{$\sim 1\%$} of the NNLO cross section).
However, the reliability of the results obtained by approximating the two-loop matrix elements in the purely soft limit may be questionable at the differential level, especially in extreme kinematic regimes.

In this paper, we provide, for the first time, fully differential predictions for the \ttH production cross section at NNLO in QCD. To overcome potential limitations of the approach used in Ref.~\cite{Catani:2022mfv},
we take one step forward and introduce a complementary approximation to estimate the two-loop $\ttH$ amplitudes.
This approximation holds in the high-energy limit for the top quarks
and is based on the so-called \textit{massification procedure} \cite{Penin:2005eh,Mitov:2006xs,Becher:2007cu,Engel:2018fsb,Wang:2023qbf}.
We show that the soft and high-energy approximations yield consistent results within their respective uncertainties. 
The two-loop contribution is ultimately obtained by combining the results from the two approximations, extending to the differential level the approach previously introduced for \ttW production in Ref.~\cite{Buonocore:2023ljm}.
Finally, we additively combine our NNLO QCD prediction with the complete set of EW corrections up to NLO\footnote{The full tower of EW corrections includes the genuine NLO EW corrections on top of the QCD Born as well as all subleading (in $\as$) terms up to NLO. More precisely, all terms of $\mathcal{O}(\as^m\alpha^n)$ with \mbox{$n+m\leq4$} are included (counting the top Yukawa coupling as $\mathcal{O}(\alpha)$).}, thus obtaining the most advanced perturbative prediction available to date both at the inclusive and differential levels.

Even if all the required amplitudes are available,
their implementation in a complete
NNLO calculation requires a method to handle and cancel IR singularities.
As in Refs.~\cite{Catani:2022mfv,Buonocore:2022pqq,Buonocore:2023ljm}, we use a process-independent implementation of the $q_T$-subtraction formalism~\cite{Catani:2007vq}, valid for the class of processes involving heavy quarks in the final state~\cite{Bonciani:2015sha,Catani:2019iny,Catani:2019hip}.
The extension of the formalism to the case in which the heavy-quark pair is accompanied by a colourless system does not pose additional complications from a conceptual point of view but requires the correct treatment of soft wide-angle QCD radiation \cite{Catani:2023tby,inprep}. 
More precisely, the requirement of back-to-back kinematics for the heavy-quark pair must be released.

The paper is organised as follows. 
Sec.~\ref{sec:double-virtual_approx} is devoted to constructing the approximations that will be used to estimate the double-virtual contribution in \ttH production: a soft Higgs boson factorisation (Sec.~\ref{subsec:soft-Higgs}) and a high-energy expansion (Sec.~\ref{subsec:massification}). 
In Sec.~\ref{sec:validation}, we carefully validate the approximations by assessing their quality at NLO, where we can compare the approximated results against the exact one-loop contribution.
We also define our procedure to combine the two approximations and conservatively assign a corresponding systematic error.
Finally, in Secs.~\ref{sec:totalXS} and \ref{sec:differential_results}, we present updated results for the total cross section and several differential distributions.
Our results are summarised in Sec.~\ref{sec:summa}.
More details on the soft-Higgs factorisation at three-loop order and on the mappings used in the high-energy expansion are provided in Appendix~\ref{app:soft} and~\ref{app:mapping}, respectively.

\section{Approximations of the double-virtual contribution}
\label{sec:double-virtual_approx}
In this paper, we consider two approximations\footnote{A different approximation treats the Higgs boson as a {\it collinear} parton radiated off a top quark. This approach was used in early NLO calculations~\cite{Dawson:1997im} of $\ttH$ production and in the more recent study of Ref.~\cite{Brancaccio:2021gcz}.} of the $\ttH$ amplitudes, each formally valid within a specific dynamical region where
\begin{enumerate}
 \item the external Higgs boson becomes soft (see Sec.~\ref{subsec:soft-Higgs});
 \item the mass of the external heavy quarks is much smaller than the typical hard scale of the process (see Sec.~\ref{subsec:massification}).
\end{enumerate}
We provide a detailed discussion of each approximation in turn.
%
%
\subsection{Soft-Higgs factorisation}\label{subsec:soft-Higgs}
We consider the hard scattering process
\begin{equation}
  \label{eq:NQQx_process}
a_1(p_1)+a_2(p_2) \!\to\! \mathcal{Q}(p_3, m)\mathcal{{\overline Q}}(p_4, m)\dots\mathcal{Q}(p_{N+1}, m)\mathcal{{\overline Q}}(p_{N+2}, m)+H(q, m_H)+F(p_F)+X(p_X)  \,,
\end{equation}
where the annihilation of the massless partons $a_1$ and $a_2$ produces
$N$ heavy quarks with (pole) mass $m$ and momenta $\{p_i\}_{i=3,\dots,N+2}$, a Higgs boson with momentum $q$
and possibly an additional system $F\,(X)$ composed of
colourless (coloured massless) particles whose total momentum $\sum p_{F_i} \,(\sum p_{X_i})$ is collectively denoted by $p_F\,(p_X)$.
We denote with $\M(\{p_i\},q)$ the renormalised all-order
scattering amplitude for the process in Eq.~(\ref{eq:NQQx_process}), where the additional dependence
on the momenta $p_F$ and $p_X$ is left understood.

We are interested in the behaviour of $\M(\{p_i\},q)$ in the limit where the Higgs boson becomes soft (i.e.\ \mbox{$q^{0}, m_H \ll Q$} with \mbox{$Q^2 = (p_1+p_2)^2$} and \mbox{$m_H \ll m$}).
In this limit, $\M(\{p_i\},q)$ fulfils the following factorisation formula,
\begin{equation}
  \label{eq:fact}
\M(\{p_i\},q)\simeq F(\as(\mu^2), \mu/m)\,\frac{m}{v} \,\left( \sum_{i=3}^{N+2} \frac{m}{p_i \cdot q} \right) \M(\{p_i\}) \,,
\end{equation}
where \mbox{$v=(\sqrt{2}G_F)^{-1/2}$} is the Higgs vacuum expectation value and $\mu$ the renormalisation scale. With $\M(\{p_i\})$ we denote the non-radiative amplitude in which the Higgs boson has been removed.
In Eq.~\eqref{eq:fact}, the symbol $\simeq$ means that we have neglected contributions that are less singular than $1/q$ in the soft-Higgs limit $q \to 0$.

The perturbative function \mbox{$F(\as(\mu^2), \mu/m)$} can be extracted by taking the soft limit of the heavy-quark scalar form factor. 
Up to ${\cal O}(\as^2)$ it reads~\cite{Bernreuther:2005gw,Ablinger:2017hst}
\begingroup
\allowdisplaybreaks
\begin{align}
  \label{eq:F}
  	F(\as(\mu^2), \mu/m) = 1 &+\frac{\as(\mu^2)}{2\pi} \left(-3C_F\right)\nonumber\\
  	&+\left(\frac{\as(\mu^2)}{2\pi}\right)^2 \!\left(\frac{33}{4}C_F^2-\frac{185}{12} C_F C_A+\frac{13}{6} C_F (n_l+n_h)-3C_F\beta_0^{(n_l)}\ln\frac{\mu^2}{m^2}\right)+{\cal O}(\as^3) \,,
  \end{align}
\endgroup
where $n_l$ is the number of massless flavours and $n_h$ the number of heavy flavours with mass $m$. 
In Eq.~(\ref{eq:F}),
\begin{equation}
	\beta^{(n_l)}_0 = \frac{11}{6}C_A - \frac{n_l}{3}
	\label{eq:first-coeff_betafunction}
\end{equation}
is the first coefficient of the QCD beta function and \mbox{$\as=\alphasNl$} is the QCD coupling renormalised in the $\MSbar$ scheme with $n_l$ active flavours%
\footnote{The result from Ref.~\cite{Ablinger:2017hst} is derived within a scheme that considers \mbox{$n_l + n_h$} active flavours. 
  For our purposes, the one-loop decoupling transformation from \mbox{$\as^{(n_l + n_h)}$} to $\alphasNl$ can be implemented by simply replacing \mbox{$\beta_0^{(n_l + n_h)}$} with $\beta_0^{(n_l)}$ in Eq.(\ref{eq:F}).}.

The factorisation formula in Eq.~\eqref{eq:fact} can be derived by using the \textit{eikonal approximation} and following the strategy applied in the soft-gluon factorisation (see, e.g.\ Ref.~\cite{Catani:2000pi}).
At the tree level, the emission of a Higgs boson
from an external leg $i$ leads to a singularity in the soft limit, where the
corresponding heavy-quark propagator \mbox{$((p_i+q)^2-m^2)^{-1}$}
becomes \mbox{$(p_i^2-m^2)^{-1}$}, with \mbox{$p_i^2=m^2$}. In contrast, the emission of a Higgs boson from an
internal heavy-quark leg will not lead to any singularity in the soft limit,
since the momentum of the emitting heavy quark is off-shell.
At the loop level, we have to consider also soft emissions from internal
lines that carry a soft momentum in the loop.
However, in the case of soft Higgs boson radiation,
the internal-line contributions, depending on the momenta of two heavy quarks, cancel out at leading power in the soft expansion.
This feature is a consequence of the fact that the Higgs boson coupling is essentially abelian
(it commutes with the soft QCD interactions in the loop)\footnote{The argument is entirely analogous to that used in the computation of the one-loop soft-gluon current in Ref.~\cite{Catani:2000pi}.}. 
Eventually, only diagrams involving a single heavy-quark momentum survive, 
and they can give a non-vanishing contribution, beyond the tree-level eikonal factorisation, to the renormalised amplitudes.
The soft limit of the heavy-quark scalar form factor controls such contribution.

The factorisation formula in Eq.~(\ref{eq:fact}) can alternatively be derived by exploiting the techniques of the \textit{low-energy theorems} (LETs)~\cite{Ellis:1975ap,Shifman:1979eb,Spira:1995rr,Kniehl:1995tn}. 
A general LET for amplitudes involving a soft Higgs boson has been successfully used to derive the effective $H\gamma\gamma$ and $Hgg$ couplings in the limit where the mass of the heavy quark in the loops is much larger than the mass of the Higgs boson. 
The theorem states that, for a generic bare amplitude $\mathcal{M}^{\mathrm{bare}}(XH)$ where $X$ denotes any state produced in association with the Higgs boson,
\begin{equation}
	\lim_{q \rightarrow 0} \mathcal{M}^{\mathrm{bare}}(XH) = \frac{m_0}{v}\frac{\partial}{\partial m_0} \mathcal{M}^{\mathrm{bare}}(X)  \,,
	\label{eq:LET}
\end{equation}
with $m_0$ being the bare mass of the fermion coupled to the Higgs boson. 
The theorem is easy to prove. 
For zero four-momentum, the kinematic derivative term in the Higgs Lagrangian can be neglected, and the (space-time independent) Higgs field can be incorporated by adding the potential energy to the bare fermion mass term, i.e.\ \mbox{$m_0 \to m_0(1+H/v)$}. In other words, the Higgs boson is treated as a background field. 

The expansion of the bare fermion propagator, for small values of \mbox{$H/v$}, is then equivalent to inserting a zero-momentum Higgs field in the non-radiative amplitude $\mathcal{M}^{\mathrm{bare}}(X)$, 
\begin{equation}
	\frac{1}{\slashed{k} -m_0} \to \frac{1}{\slashed{k} -m_0}\frac{m_0}{v}\frac{1}{\slashed{k} -m_0} \,,
\end{equation}
and thus generating the amplitude $\mathcal{M}^{\mathrm{bare}}(XH)$. 
The renormalisation of the bare quantities is performed after evaluating the right-hand side of Eq.~\eqref{eq:LET}.
The caveat of the theorem is that the differentiation in Eq.~\eqref{eq:LET} only acts on the bare masses appearing in the propagators of the massive fermions. In contrast, bare mass-dependent couplings must be treated as constants.

Having Eq.~\eqref{eq:LET} in mind, we can now go back to our original discussion on the derivation of the perturbative function \mbox{$F(\as(\mu^2), \mu/m)$} in Eq.~\eqref{eq:F} via the LET. 
The basic observation is that, at the bare level, we have
\begin{equation}
	\lim_{q \rightarrow 0} \M^{\mathrm{bare}}(\{p_i\},q) = \frac{m_{0}}{v} \left( \sum_{i=3}^{N+2}\,\frac{m_{0}}{p_i \cdot q} \right) \M^{\mathrm{bare}}(\{p_i\}) \,,
\end{equation}
valid at all orders in perturbative QCD.
The renormalisation of the heavy-quark mass and its wave function induces a modification of the Higgs coupling to the heavy quark $\mathcal{Q}$.
This can be seen as an effective coupling $\delta_{H \mathcal{Q}\bar{\mathcal{Q}} }$ in the soft-Higgs limit, with 
\begin{equation}
	\delta_{H \mathcal{Q}\bar{\mathcal{Q}} } = F(\as(\mu^2), \mu/m) -1 \,.
	\label{eq:deltaHQQ-F}
\end{equation}
To prove the validity of Eq.~\eqref{eq:deltaHQQ-F} up to ${\cal O}(\as^2)$, we need to start from the bare amplitude for the emission of a soft Higgs boson from a heavy quark $\mathcal{Q}$ with momentum $p$. 
By exploiting Eq.~\eqref{eq:LET} we can write 
\begin{equation}
	\lim_{q \rightarrow 0} \M_{\mathcal{Q} \to \mathcal{Q} H}^{\mathrm{bare}}(p,q) = \frac{m_{0}}{v} \frac{\partial}{\partial m_{0}} \M_{\mathcal{Q} \to \mathcal{Q}}^{\mathrm{bare}}(p) \biggl |_{p^2 = m^2} \,,
	\label{eq:LET_selfenergy}
\end{equation}
where 
\begin{equation}
	\M_{\mathcal{Q} \to \mathcal{Q}}^{\mathrm{bare}}(p) = \overline{\mathcal{Q}}_{0}(p) \left( -m_{0} + \Sigma(p,m_0) \right) \mathcal{Q}_{0}(p)	\,.
\end{equation}
We emphasise that it is crucial to evaluate the right-hand side of Eq.~\eqref{eq:LET_selfenergy} in the limit \mbox{$p^2 \to m^2$} only after taking the derivative with respect to the bare mass, as this ensures the cancellation of all gauge-dependent terms.

It is now clear that evaluating Eq.~\eqref{eq:LET_selfenergy} requires the knowledge of the ${\cal O}(\as^2)$ contribution to the unrenormalised heavy-quark self-energy
\begin{equation}
	\Sigma(p,m_0) = m_{0}\Sigma_S(p^2,m_0) + \slashed{p}\Sigma_V(p^2,m_0) \,.
\end{equation}
By following the conventions of Refs.~\cite{Broadhurst:1991fy, Gray:1990yh}, we decompose the scalar and vector parts of the quark self-energy as
\begingroup
\allowdisplaybreaks
\vspace{-0.3cm}
\begin{align}
	\Sigma_S(p^2,m_0) &=  - \sum_{n=1}^{+\infty}\,\biggl[ \frac{g_0^2}{(4\pi)^{d/2}(p^2)^{\epsilon}} \biggr]^n \bigl( A_n(m^2_0/p^2) -  B_n(m^2_0/p^2) \bigr) \,,\\
	\Sigma_V(p^2,m_0) &=  - \sum_{n=1}^{+\infty}\,\biggl[ \frac{g_0^2}{(4\pi)^{d/2}(p^2)^{\epsilon}} \biggr]^n \,B_n(m^2_0/p^2) \,,
	\label{eq:sigmaSsigmaV}
\end{align}
\endgroup
where $g_0$ denotes the bare (dimensionful) coupling constant and \mbox{$d=4 -2\epsilon$} is the number of space-time dimensions. 
The functions \mbox{$A_n(m^2_0/p^2)$} and \mbox{$B_n(m^2_0/p^2)$} are dimensionless and depend on $d$, the gauge parameter and the ratio \mbox{$m^2_0/p^2$}.

Having set the notation, we can compute the right-hand side of Eq.~\eqref{eq:LET_selfenergy} as
\begingroup
\allowdisplaybreaks
\begin{align}
	m_{0} \frac{\partial}{\partial m_{0}} \M_{\mathcal{Q} \to \mathcal{Q}}^{\mathrm{bare}}(p) \biggl |_{p^2 = m^2} \!\!\!\!\! &= 
	-m_{0}  \overline{\mathcal{Q}}_{0} \mathcal{Q}_{0} \biggl \{ 1 
	 +  \frac{g_0^2}{(4\pi)^{d/2} m^{2\epsilon}}\biggl[ A_1(Z_m^2) - B_1(Z_m^2)  
	 + 2 Z_m^2 A^{'}_1(Z_m^2) \notag \\
	 &\hspace*{14em}+2 Z_m^2 (Z_m^{-1} -1)B^{'}_1(Z_m^2) \biggr] \notag \\
	 &+\left( \frac{g_0^2}{(4\pi)^{d/2} m^{2\epsilon}} \right)^2\biggl[  A_2(Z_m^2) - B_2(Z_m^2) + 2 A^{'}_2(Z_m^2) \biggr]  + \mathcal{O}(g_0^6)
	 \biggr\} \,,
\end{align}
\endgroup
where $A_n^{'} (B_n^{'})$ are the derivatives of $A_n (B_n)$ with respect to \mbox{$m^2_0/p^2$}, while
\begin{equation}
	Z_m = \frac{m_0}{m} = 1 + \sum_{n=1}^{+\infty}\,\biggl[\frac{g_0^2}{(4\pi)^{d/2} m^{2\epsilon}} \biggr]^n \,M_n
	\label{eq:Z_m}
\end{equation}
is the mass renormalisation function in the on-shell scheme.
Explicit expressions of the coefficients $M_n$, up to the two-loop order, can be extracted from Ref.~\cite{Broadhurst:1991fy}.

In order to exploit the results of Refs.~\cite{Broadhurst:1991fy, Gray:1990yh}, we need to expand the functions $A_n(Z_m^2)$ and $B_n(Z_m^2)$ around \mbox{$Z_m^2 = 1$}.
After performing this expansion and neglecting contributions of $\mathcal{O}(g_0^6)$, we obtain 
\begin{align}
	\hspace{-0.7cm} 
	m_{0} \frac{\partial}{\partial m_{0}} \M_{\mathcal{Q} \to \mathcal{Q}}^{\mathrm{bare}}(p)\biggl |_{p^2 = m^2} &= 
	-m_{0}  \overline{\mathcal{Q}}_{0} \mathcal{Q}_{0} \biggl \{ 1 +  \frac{g_0^2 }{(4\pi)^{d/2} m^{2\epsilon}}\biggl[ A_1 - B_1 + 2 A^{'}_1 \biggr]  \notag \\
	&+ \left(\frac{g_0^2}{(4\pi)^{d/2}m^{2\epsilon}} \right)^2 \biggl[  A_2 - B_2 + 2 A^{'}_2  + M_1 ( 6 A^{'}_1 - 4 B^{'}_1 + 4 A^{''}_1 )\biggr] 
	+ \mathcal{O}(g_0^6)
	 \biggr\} \,,
	 \label{eq:bare_amplitude}
\end{align}
which is a gauge-independent result. 
This statement can be easily proven by exploiting the relations among the functions $A_n$, $B_n$ and $M_n$, derived in Ref.~\cite{Broadhurst:1991fy}.
The $A_n$, $B_n$ functions and their derivatives without explicit functional dependence in Eq.~\eqref{eq:bare_amplitude} have to be understood as evaluated at \mbox{$Z_m^2=1$}.

The next step involves the on-shell renormalisation of the heavy-quark mass and wave function,
\begin{equation}
	m_{0}  \overline{\mathcal{Q}}_{0} \mathcal{Q}_{0} = m  \overline{\mathcal{Q}} \mathcal{Q} \,Z_m Z_2 \,,
\end{equation}
where $Z_2$ is the residue at the pole mass of the heavy-quark Feynman propagator.
$Z_2$ can be expanded as
\begin{equation}
	Z_2 = 1 + \sum_{n=1}^{+\infty}\,\biggl[ \frac{g_0^2}{(4\pi)^{d/2}m^{2\epsilon}} \biggr]^n \,F_n \,,
	\label{eq:Z_2}
\end{equation}
where explicit expressions of the functions $F_n$, up to two-loop order, are provided in Ref.~\cite{Broadhurst:1991fy}.

Finally, we renormalise the strong coupling in the $\MSbar$ scheme via
\begin{equation}
	\frac{g_0^2}{4\pi} = \biggl(\frac{e^{\gamma_E} \mu^2}{4\pi} \biggr)^{\!\epsilon} \alphasNf(\mu^2) \biggl[ 1- \frac{\alphasNf(\mu^2) }{2 \pi } \frac{\beta_0^{(n_f)}}{\epsilon}  + \mathcal{O}(\as^2) \biggr] \,,
	\label{eq:alpha_renorm}
\end{equation}
with \mbox{$n_f = n_l + n_h$}, and apply the one-loop decoupling relation (see e.g.\ Ref.~\cite{Steinhauser:2002rq} for a review)
\begingroup
\allowdisplaybreaks
\begin{equation}
	\alphasNf(\mu^2) = \alphasNl(\mu^2) \biggl[ 1 +  \frac{\alphasNl(\mu^2)}{2 \pi}  \frac{n_h}{3}\biggl( \Gamma(\epsilon) e^{\gamma_E \epsilon} \biggl( \frac{\mu^2}{m^2} \biggr)^{\!\epsilon} -\frac{1}{\epsilon} \biggl) + \mathcal{O}(\as^2)   \biggr]  \,,
	\label{oneloop_decoupling}
\end{equation}
\endgroup
since we want to work in the decoupling scheme
where the $n_h$ heavy quarks of mass $m$ do not contribute to the running of $\as$.
The correction, up to ${\cal O}(\as^2)$, to the Higgs--quark coupling in the soft-Higgs limit is ultimately given by
\begingroup
\allowdisplaybreaks
\begin{align}
	\delta_{H \mathcal{Q}\bar{\mathcal{Q}} } &= \frac{\alphasNl(\mu^2)}{2\pi} \left(-3C_F\right)  \notag \\
 	&+\left(\frac{\alphasNl(\mu^2)}{2\pi}\right)^{\!2}\left(\frac{33}{4}C_F^2-\frac{185}{12} C_F C_A+\frac{13}{6} C_F (n_l+n_h)-3C_F\beta_0^{(n_l)}\ln\frac{\mu^2}{m^2}\right)+{\cal O}(\as^3) \,,
\end{align}
\endgroup
consistent with Eq.~\eqref{eq:F}.
This derivation proves that, up to the two-loop order in QCD, the result from applying the Higgs LET coincides with the soft limit of the massive scalar form factor.
In Appendix \ref{app:soft}, we show how this result can be extended to the three-loop order.

In the following, we will apply the soft-factorisation formula to construct an approximation of the two-loop $t{\bar t}H$ amplitudes. 
This can be achieved by setting \mbox{$N=2$} in Eq.~\eqref{eq:fact} and relying on the two-loop amplitudes for $t\bar{t}$ production~\cite{Barnreuther:2013qvf}.
The factorisation formula can also provide a nontrivial check of future multi-loop computations for processes involving the Higgs boson and heavy quarks in the soft-Higgs limit.

\subsection{Mass factorisation in the high-energy limit}\label{subsec:massification}
We now consider the \mbox{$2 \to n$} scattering process
\begin{equation}
	a_1(p_1) + a_2(p_2) \to a_3(p_3) + \dots + a_{n+2}(p_{n+2}) + F\,,
	\label{eq:2ton_process}
\end{equation}
where the collision of two partons with flavours $a_1, a_2$ and momenta $p_1, p_2$ produces $n$ final-state partons with flavours $a_i$ and momenta $p_i$ (\mbox{$i=3,... n+2$}), plus a possible system $F$ of colourless particles.

The corresponding scattering amplitude, $\M$, is treated as a vector in colour space,
\begin{equation}
	\M\left( \{ p_i\}, \frac{Q^2}{\mu^2}, \as(\mu^2), \epsilon \right) \equiv \ket{\M} \,,
\end{equation}
where $\mu$ is the renormalisation scale and $Q$ the overall hard-scattering scale, which we may choose as the invariant mass of the event, i.e.\ \mbox{$Q^2 = (p_1+p_2)^2$}.

The IR structure of the amplitude will take a completely different form depending on the massive or massless nature of the final-state particles.
Indeed, amplitudes with massless partons exhibit poles related to collinear singularities, whereas in the massive case collinear divergences are screened by the mass $m$. This instead gives rise to logarithmically enhanced contributions of Sudakov type~\cite{Sudakov:1954sw}, often called \textit{quasi-collinear} singularities~\cite{Catani:2000ef}.
The predictable nature of IR singularities allows us, through the so-called \textit{massification} procedure, to relate the leading logarithms, $\ln(m^2/Q^2)$, arising in the massive case to the pole structure emerging in the corresponding massless limit.
This correspondence between massive and massless amplitudes holds up to power corrections in $m$ and is thus particularly effective, in the small-mass case, to approximate massive amplitudes starting from their massless counterpart.

The massification technique was first applied to the computation of NNLO QED corrections to large-angle Bhabha scattering~\cite{Penin:2005eh,Becher:2007cu} and later extended to any \mbox{$2 \to n$} scattering process~\cite{Mitov:2006xs} in QCD. This extension is based on the factorisation properties of scattering amplitudes in the IR singular limits, a principle that also holds true in non-abelian gauge theories. Recently, the technique has garnered additional interest, with further advancements made by incorporating the contributions from massive fermionic loops \cite{Wang:2023qbf}.
Next-to-leading power corrections were also investigated for specific processes (see e.g.\ Refs.~\cite{Penin:2014msa, Penin:2016wiw, Liu:2017vkm} where the all-order result for the leading mass-suppressed terms in the double-logarithmic approximation has been derived). Still, a general treatment has yet to be formulated.

In our context, we will exploit the massification procedure to formulate a high-energy approximation for the \ttH two-loop amplitudes, representing the cutting-edge of current computational methods. 
A similar approach has been adopted in Refs.~\cite{Buonocore:2022pqq,Buonocore:2023ljm,Mazzitelli:2024ura} to approximate the two-loop massive amplitudes for the $b \bar b W$, $t \bar t W$ and $b \bar b Z$ processes.

\subsubsection{Factorisation of QCD amplitudes}
The massification technique relies on the factorisation properties of QCD amplitudes~\cite{Catani:1998bh,Sterman:2002qn}.
For the process in Eq.~\eqref{eq:2ton_process}, if only massless external partons are involved, the general form of the factorised amplitude is 
\begin{equation}
	\ket{\M} =  \J_0\left( \frac{Q'^2}{\mu^2}, \as(\mu^2), \epsilon \right) \Soft_{\hspace{-0.1cm}0}\left( \{ p_i\}, \frac{Q'^2}{\mu^2}, \frac{Q'^2}{Q^2}, \as(\mu^2), \epsilon \right) \ket{\Hard} \,,
	\label{eq:massless_QCD_factorisation}
\end{equation}
where $\ket{\Hard}$ is the process-dependent hard function, which describes the short-distance dynamics of the hard scattering and satisfies a colour decomposition analogous to the amplitude $\ket{\M}$, while $\Soft_{\hspace{-0.1cm}0}$ is the soft function, encoding the wide-angle soft radiation arising from the overall colour flow, expressed as an operator in colour space. Finally, $\J_0$ is the jet function, which encapsulates collinear-sensitive contributions and can be written in a factorised form, 
\begin{equation}
	\J_0\left( \frac{Q'^2}{\mu^2}, \as(\mu^2), \epsilon \right) = \prod_{i=1}^{n+2} \J_0^{[i]}\left( \frac{Q'^2}{\mu^2}, \as(\mu^2), \epsilon \right) \,,
\end{equation} 
with respect to each external parton $a_i$.
The construction of the soft and jet functions requires the specification of at least one independent momentum scale, $Q'$, which represents the factorisation scale. Such a scale, distinct from $Q$ and $\mu$, may play a role in the case of a multi-scale process, where strong hierarchies between invariants can arise. In the following, the factorisation scale is chosen to be equal to the hard scale $Q$. 

The predictive power of Eq.~\eqref{eq:massless_QCD_factorisation} depends on the properties of jet, soft and hard functions.
The jet function $\J_0$ collects all poles in dimensional regularisation, which arise from the overlap of soft and collinear enhancements in perturbation theory. The colour matrix $\Soft_{\hspace{-0.1cm}0}$ provides, at most, a single soft pole per loop, and although it depends on the process kinematics, it is entirely determined by a matrix of soft anomalous dimensions. The hard function $\Hard$ is constructed to be IR finite. 
Therefore, the factorisation in Eq.~\eqref{eq:massless_QCD_factorisation} is uniquely defined only up to finite contributions in the various functions. 
For example, we can shift IR finite terms between the jet and hard function and/or process-independent soft contributions (proportional to the identity matrix in the colour basis space) between the soft matrix and the jet function. It is thus relevant to fix a scheme for defining the various functions in Eq.~\eqref{eq:massless_QCD_factorisation}.
Following Ref.~\cite{Sterman:2002qn}, a convenient explicit expression for the jet functions can be obtained by applying Eq.~\eqref{eq:massless_QCD_factorisation} to the simplest amplitudes for which the factorisation formula holds, i.e.\ the gauge-invariant quark and gluon form factors. 
We thus define our jet functions by the relation
\begingroup
\allowdisplaybreaks
\begin{equation}
	\J_0^{[i]}\left( \frac{Q^2}{\mu^2}, \as(\mu^2), \epsilon \right) = \J_0^{[\bar i\,]}\left( \frac{Q^2}{\mu^2}, \as(\mu^2), \epsilon \right) = \biggl( \mathcal{F}_0^{[i \bar i \to F]}\left( \frac{Q^2}{\mu^2}, \as(\mu^2), \epsilon \right)  \biggr)^{1/2}  \,,~~~~ i \in \{q,g\} \,,
	\label{eq:massless_jet_function}
\end{equation}
\endgroup
where \mbox{$\mathcal{F}_0^{[i \bar i \to F]}$} (\mbox{$i=q,g$}) are the quark and gluon {\it timelike} form factors.
The asymptotic behaviour of the form factors at high energy\footnote{In this limit, the form factor is often called \textit{Sudakov form factor} since it exhibits a doubly logarithmic behaviour, as a consequence of its soft and collinear singularities.} has been the subject of theoretical studies for almost half a century. 
The all-order exponentiation has been extensively studied in the literature (see, e.g.\ Refs.~\cite{Collins:1989bt, Magnea:2000ss, Magnea:2000ep}), while fixed-order calculations of the massless form factors are available up to four-loop order in QCD~\cite{Lee:2022nhh}.
We note that our choice of the jet function~\eqref{eq:massless_jet_function} corresponds to a particular scheme for the form factors, in which the hard and soft functions are set to unity, \mbox{$\Soft_{\hspace{-0.1cm}0}= 1$} and \mbox{$\Hard= 1$}, in Eq.~\eqref{eq:massless_QCD_factorisation}.
Having set the scheme for the jet function, the soft function turns out to be independent of collinear dynamics and can be constructed from eikonal amplitudes, as described in detail in Ref.~\cite{Aybat:2006mz}.

We are now ready to consider the situation in which one or more external particles acquire a non-vanishing mass $m$, much smaller than the characteristic hard scale $Q$, i.e.\ \mbox{$m \ll Q$}.
In this limit, the massive amplitude can be written, up to power-suppressed contributions in \mbox{$m/Q$}, in a factorised form similar to Eq.~\eqref{eq:massless_QCD_factorisation},
\begin{equation}
	\ket{\M_m} =  \J\left( \frac{Q^2}{\mu^2}, \frac{m^2}{\mu^2}, \as(\mu^2), \epsilon \right) \Soft\left( \{ p_i\}, \frac{Q^2}{\mu^2}, \frac{m^2}{\mu^2}, \as(\mu^2), \epsilon \right) \ket{\Hard} \,,
	\label{eq:massive_QCD_factorisation}
\end{equation}
where the nontrivial mass dependence only appears in the jet and soft functions. Since the massive hard function is insensitive to the collinear dynamics, it is affected by power corrections in \mbox{$m/Q$} and thus, in Eq.~\eqref{eq:massive_QCD_factorisation}, can be identified with its corresponding massless counterpart. The jet function $\J$ encodes all quasi-collinear contributions from the external partons and, similarly to the massless case, can be written in the form 
\begin{equation}
	\J\left( \frac{Q^2}{\mu^2}, \frac{m^2}{\mu^2}, \as(\mu^2), \epsilon \right) = \prod_{i=1}^{n+2} \J^{[i]}\left( \frac{Q^2}{\mu^2}, \frac{m^2}{\mu^2}, \as(\mu^2), \epsilon \right) \,.
\end{equation} 
Even with final-state masses, Eq.~\eqref{eq:massive_QCD_factorisation} retains a scheme-choice ambiguity due to the freedom in constructing the jet and soft functions. As in the massless case, we resolve this ambiguity by requiring that
\begin{equation}
	\J^{[i]}\left( \frac{Q^2}{\mu^2}, \frac{m^2}{\mu^2}, \as(\mu^2), \epsilon \right) = \biggl( \mathcal{F}^{[i \bar i \to F]}\left( \frac{Q^2}{\mu^2}, \frac{m^2}{\mu^2}, \as(\mu^2), \epsilon \right)  \biggr)^{\!1/2}  \,,~~~~ i \in \{q,g,\mathcal{Q}\} \,,
	\label{eq:massive_jet_function}
\end{equation}
where \mbox{$\mathcal{F}^{[\mathcal{Q}\overline{\mathcal{Q}} \to F]}$} is the heavy-quark form factor, while \mbox{$\mathcal{F}^{[q\bar q \to F]}$} and \mbox{$\mathcal{F}^{[gg \to F]}$} coincide with the massless form factors, \mbox{$\mathcal{F}_0^{[q\bar q \to F]}$} and \mbox{$\mathcal{F}_0^{[gg \to F]}$}, up to \mbox{$\ln(m/\mu)$} and finite terms due to vertex corrections involving heavy-quark loops.

We note that if we neglect contributions from internal heavy-quark loops, the soft function will be only affected by power-suppressed contributions in \mbox{$m/Q$}.
Indeed, logarithmically enhanced terms in the mass have a collinear origin and, thus, they can be reabsorbed in the definition of the jet function \mbox{$\J^{[\mathcal{Q}]}$}. This is precisely what happens in the scheme where the jet function \mbox{$\J^{[\mathcal{Q}]}$} is defined to be the square root of the massive form factor \mbox{$\mathcal{F}^{[\mathcal{Q} \overline{\mathcal{Q}} \to F]}$}.
In this case, as in Ref.~\cite{Mitov:2006xs}, the soft and hard functions are identical in both the massless and massive cases, provided that \mbox{$m \ll Q$}. This implies that the factorisation formula in Eq.~\eqref{eq:massive_QCD_factorisation} is also valid for the massive form factor \mbox{$\mathcal{F}^{[\mathcal{Q} \overline{\mathcal{Q}} \to F]}$}, by setting \mbox{$\Soft = 1$} and \mbox{$\Hard = 1$}.

As a first approximation, we can neglect the contribution from diagrams involving heavy-quark loops. 
In this way, we focus only on mass effects related to the external legs. The differences between the massless and massive schemes are entirely encoded in the jet functions. 
This observation leads to the remarkably simple and direct relation~\cite{Mitov:2006xs} between massless and massive UV renormalised amplitudes for the process in Eq.~(\ref{eq:2ton_process}), 
\begin{equation}
	\ket{ \M_m } =
	\left( Z^{(m|0)}_{[\mathcal{Q}]}\left(\alphasNl, \frac{\mu^2}{m^2}, \epsilon \right) \right)^{\!n_{\mathcal{Q}}/2} \ket{ \M} \,,
	\label{eq:Moch_massification}
\end{equation}
where $n_{\mathcal{Q}}$ denotes the number of external quarks we want to promote from massless to massive particles of mass $m$. 
The function \mbox{$Z^{(m|0)}_{[\mathcal{Q}]}$} is universal, process-independent and can be viewed as a factor connecting two different regularisation schemes. It is given in terms of the ratio between the massive and massless quark form factors,
\begin{align}
	Z^{(m|0)}_{[\mathcal{Q}]}\left(\alphasNl, \frac{\mu^2}{m^2}, \epsilon \right) 
	=  \mathcal{F}^{[\mathcal{Q} \overline{\mathcal{Q}} \to F]}\left( \frac{Q^2}{\mu^2}, \frac{m^2}{\mu^2}, \alphasNl(\mu^2), \epsilon \right)\,  \biggl( \mathcal{F}_0^{[q \bar q \to F]}\left( \frac{Q^2}{\mu^2}, \alphasNl(\mu^2), \epsilon \right)  \biggr)^{-1}  \,,
	\label{eq:ZMoch_functions}
\end{align}
and an explicit expression, up to two-loop order, can be found in Ref.~\cite{Mitov:2006xs}. The process independence of \mbox{$Z^{(m|0)}_{[\mathcal{Q}]}$} is manifest: the dependence on the hard scale $Q$ cancels in the ratio, and only process-independent energy scales $\mu, m$ are left.
A final remark is that, in Eq.~\eqref{eq:Moch_massification}, the mass is always chosen to be the pole mass, and the strong coupling $\as$ is renormalised in the $\MSbar$ scheme with $n_l$ massless active flavours.

%
We now discuss how Eq.~\eqref{eq:Moch_massification} is affected by the inclusion of heavy-quark loop effects.
These terms were computed for the first time in Ref.~\cite{Becher:2007cu} in the context of two-loop QED corrections for Bhabha scattering. There, it was shown that a non-trivial soft function arises once vacuum polarisation diagrams with massive fermions are considered. 
In the context of heavy-to-light form factors, a similar factorisation formula has been presented in Ref.~\cite{Engel:2018fsb}, where previous NNLO analyses have been extended to the case of two different external masses satisfying the hierarchy \mbox{$M \gg m$}. 
In the case of heavy-quark production, a direct computation of these process-dependent contributions was carried out in Refs.~\cite{Czakon:2007ej,Czakon:2007wk} for the $q \bar q$ and $gg$ partonic channels, respectively.

Very recently, a general factorisation formula accounting for the above contributions was presented in Ref.~\cite{Wang:2023qbf}.
It can be applied to any scattering amplitude\footnote{The only caveat, as pointed out by the authors of Ref.~\cite{Wang:2023qbf}, is that all parton masses must be much smaller than the characteristic hard scale of the process.}, in the high-energy limit, up to the two-loop order. 
Therefore, Eq.~\eqref{eq:Moch_massification} must be replaced by 
\begin{equation}
	\ket{ \M_m } =
	\prod_i \left( Z^{(m|0)}_{[i]}\left(\alphasNf, \frac{\mu^2}{m^2}, \epsilon \right) \right)^{\!1/2} \SSoft \left(\alphasNf,\frac{\mu^2}{s_{ij}},\frac{\mu^2}{m^2}, \epsilon \right) \ket{ \M} \,,
	\label{eq:generalised_massification}
\end{equation}
where $i$ runs over all external partons $\{a_i\}$.
The strong coupling $\as$ is renormalised with \mbox{$n_f = n_l + n_h$} flavours, where $n_l$ ($n_h$) denotes the number of light (heavy) quarks in the theory.

Analogously to Eq.~\eqref{eq:ZMoch_functions}, the $Z$-factors \mbox{$Z^{(m|0)}_{[i]}$} are defined as the ratio of the massive (\mbox{$\mathcal{F}^{[i \bar i \to F]}$}) and massless (\mbox{$\mathcal{F}_0^{[i \bar i \to F]}$}) form factors. They can be expanded as\footnote{In order to match the notation and conventions of Refs.~\cite{Mitov:2006xs,Wang:2023qbf}, in this section we use an expansion in powers of $\as/(4\pi)$.}
\begin{equation}
	Z_{[i]}^{(m|0)}\left(\alphasNf, \frac{\mu^2}{m^2}, \epsilon \right) = 1+ \sum_{n=1}^{\infty} \left( \frac{\alphasNf(\mu^2)}{4 \pi} \right)^{\!n} \!Z_{[i]}^{(n)}\left( \frac{\mu^2}{m^2},\epsilon \right)\,, ~~~~ i \in \{q,g,\mathcal{Q}\} \,,
\end{equation}
and now contain $n_h$-dependent contributions up to $n_h^2$. 
The $Z$-factor for a massless quark starts contributing at $\mathcal{O}(\as^2)$ (i.e.\ \mbox{$Z_{[q]}^{(1)} = 0$}), while the gluon function \mbox{$Z_{[g]}^{(m|0)}$} receives $n_h$ contributions already at one-loop level, due to the one-loop correction to the gluon form factor coming from a heavy-quark bubble or triangle.
It is worth mentioning that the function \mbox{$Z_{[g]}$}, up to two-loop order, is equal to the on-shell wave-function renormalisation constant of a gluon, evaluated in a physical gauge. 
The explicit expression of \mbox{$Z_{[i]}^{(m|0)}$} can be found in Ref.~\cite{Wang:2023qbf}. 

The soft function $\SSoft$ in Eq.~\eqref{eq:generalised_massification} is given by
\begin{equation}
	\SSoft \left(\alphasNf,\frac{\mu^2}{s_{ij}},\frac{\mu^2}{m^2}, \epsilon \right) = 1+ \left( \frac{\alphasNf(\mu^2)}{4 \pi} \right)^{\!2} n_h \sum_{i>j} (-\mathbf{T}_i \cdot \mathbf{T}_j) \,S^{(2)}\left(\frac{\mu^2}{s_{ij}},\frac{\mu^2}{m^2},\epsilon \right)  + \mathcal{O}(\as^3) \,,
	\label{two-loop_soft-function}
\end{equation}
where $n_h$ is the number of flavours with mass $m$ present in the theory\footnote{In our discussion, we consider only the case in which all massive quarks have the same mass $m$. For a more general treatment, we refer the reader to Ref.~\cite{Wang:2023qbf} where effects due to a quark of mass \mbox{$m_h \sim m$}, \mbox{$m_h \ne m$} are also accounted for.}, and 
\begin{equation}
	s_{ij} = 2 \sigma_{ij} p_i \cdot p_j + i0 \,,
	\label{eq:sij_inv}
\end{equation}
with \mbox{$\sigma_{ij} = +1$} if the momenta $p_i$ and $p_j$ are both incoming or outgoing and \mbox{$\sigma_{ij} = -1$} otherwise.
The two-loop coefficient $S^{(2)}$ reads \cite{Wang:2023qbf}
\begin{equation}
	S^{(2)}\left(\frac{\mu^2}{s_{ij}},\frac{\mu^2}{m^2},\epsilon \right) = T_R \left( \frac{\mu^2}{m^2} \right)^{\!2\epsilon} \!\left(-\frac{4}{3\epsilon^2} + \frac{20}{9 \epsilon} - \frac{112}{27} - \frac{4 \zeta_2}{3} \right) \ln\left(\frac{-s_{ij}}{m^2}\right)\, .
	\label{two-loop_soft-coeff}
\end{equation}
We note that the coefficient $S^{(2)}$ is completely analogous to that computed in Ref.~\cite{Becher:2007cu} for QED corrections to Bhabha scattering, with the only difference that the overall factor \mbox{$n_h \alpha^2$} must be substituted with \mbox{$n_h T_R C_F \as^2$}.

\subsubsection{Massive amplitudes for \texorpdfstring{$\mathcal{Q}\overline{\mathcal{Q}} H$}{QQH} production}\label{subsubsec:massification_HQQ}
We now focus on the $\mathcal{Q}\overline{\mathcal{Q}}H$ process,
\begin{equation}
	c(p_1) + \bar{c}(p_2) \to \mathcal{Q}(p_3, m) + \overline{\mathcal{Q}}(p_4, m) + H(q, m_H)\,,~~~~~ c \in \{q,g\} \,,
	\label{HQQ_process}
\end{equation}
and apply the mass factorisation formula in Eq.~\eqref{eq:generalised_massification} to approximate the corresponding two-loop amplitudes\footnote{During the completion of our work, a similar procedure was presented in Ref.~\cite{Wang:2024pmv} and applied to a set of fully massless amplitudes, thus not including the dependence on the Higgs boson mass.}.
These approximated amplitudes not only provide an estimate of the double-virtual contribution to \ttH production in the boosted regime, but might also be exploited to complete an NNLO computation for $b \bar b H$ production in the four-flavour scheme.

The starting point of our construction is represented by the massless two-loop amplitudes analytically computed, in leading-colour approximation, in Ref.~\cite{Badger:2021ega} for the production of a Higgs boson in association with a pair of (massless) bottom quarks.
Some comments on such massless amplitudes are in order.
First, the bottom quark is treated as massless, while the corresponding Yukawa coupling is kept finite. This approach is feasible because a mixed renormalisation scheme can be employed, provided that only pure QCD corrections are considered. In this scheme, the bottom-quark mass and wave function are renormalised on-shell, while the Yukawa and strong couplings are (renormalised) in the $\overline{\text{MS}}$ scheme.
Secondly, diagrams involving heavy top-quark loops are not included, while massless bottom-quark loops attached to a Higgs boson give rise to a vanishing contribution due to an odd number of Dirac matrices in the trace.
Finally, the one-loop and two-loop finite remainders are provided in the IR subtraction scheme of Ref.~\cite{Catani:1998bh}.

To exploit the results of Refs.~\cite{Catani:2023tby,inprep} on the soft-parton contributions at low transverse momentum, which are required in the $q_T$-subtraction formalism for $\ttH$ production, it is convenient for us to define the finite remainder of the massive two-loop amplitude in the scheme of Refs.~\cite{Becher:2009cu,Becher:2009qa,Ferroglia:2009ii} (see also Ref.~\cite{Gardi:2009qi}).
This requires a conversion of the massless finite remainders of Ref.~\cite{Badger:2021ega}, before applying the mass factorisation formula in Eq.~\eqref{eq:generalised_massification}.
Additionally, we have to take care of mass logarithms, \mbox{$l_{\mu m} = \ln\frac{\mu^2}{m^2}$}, arising from the change of the Yukawa renormalisation from the $\MSbar$ scheme used in Ref.~\cite{Badger:2021ega} to the on-shell scheme applied in this work. 

Having set our conventions, we can directly connect the (all-order) massive finite remainder at scale $\mu$ to the corresponding massless one in the same IR subtraction scheme via 
\begingroup
\allowdisplaybreaks
\begin{align}
	 \ket{ \M_{m}^{\finSCET} } &= \ZNeubert_{m \ll \mu_h}^{-1}\left(\alphasNf,\frac{\mu^2}{s_{ij}},\frac{\mu^2}{m^2}, \epsilon\right) Z_{[\mathcal{Q}]}^{(m|0)}\left(\alphasNf,\frac{\mu^2}{m^2}, \epsilon \right) Z_{[c]}^{(m|0)}\left(\alphasNf,\frac{\mu^2}{m^2}, \epsilon \right) \notag \\
	&\times \SSoft\left(\alphasNf,\frac{\mu^2}{s_{ij}},\frac{\mu^2}{m^2} , \epsilon\right) \ZNeubert_{(m=0)}\left(\alphasNf, \frac{\mu^2}{s_{ij}}, \epsilon\right) \,\ket{ \M_{(m=0)}^{\finSCET} } + \mathcal{O}\left( \frac{m}{\mu_h}\right) \,,
	\label{master_equation4}
\end{align}
\endgroup
where $\mu_h$ represents any hard scale involved in the process and \mbox{$c \in \{q, g\}$} stands for the partonic channel in Eq.~\eqref{HQQ_process}.
The massless finite remainder \mbox{$\ket{ \M_{(m=0)}^{\finSCET} }$} is defined to be evaluated with \mbox{$n_f = n_l + n_h$} massless flavours.

In the previous formula, \mbox{$\ZNeubert_{(m=0)}$} is the massless subtraction operator of Refs.~\cite{Becher:2009cu,Becher:2009qa}, while \mbox{$\ZNeubert_{m \ll \mu_h}$} denotes the operator that subtracts the poles of the corresponding massive amplitude, expanded in the small-mass limit (\mbox{$m \ll \mu_h$}).
All IR poles in $\epsilon$ cancel, and the r.h.s.\ of Eq.~\eqref{master_equation4} can be rewritten as
\begin{equation}
	\ket{ \M_{m}^{\finSCET} } = \boldsymbol{\mathcal{F}}_{[c]}\left(\alphasNf,\frac{\mu^2}{m^2}, \frac{\mu^2}{s_{ij}} \right)\,\ket{ \M_{(m=0)}^{\finSCET} } + \mathcal{O}\left( \frac{m}{\mu_h}\right) \,,
	\label{master_equation5}
\end{equation}
where \mbox{$\boldsymbol{\mathcal{F}}_{[c]}$} is an operator in colour space, depending on the partonic channel $c$. 
It can be expanded in powers of the strong coupling as
\begin{equation}
	\boldsymbol{\mathcal{F}}_{[c]} \left(\alphasNf,\frac{\mu^2}{m^2}, \frac{\mu^2}{s_{ij}} \right) = 
	1 + \frac{\alphasNf(\mu^2)}{4\pi} \boldsymbol{\mathcal{F}}_{[c]}^{(1)}\left(\frac{\mu^2}{m^2}, \frac{\mu^2}{s_{ij}} \right) + \left(\frac{\alphasNf(\mu^2)}{4\pi}\right)^{\!2} \boldsymbol{\mathcal{F}}_{[c]}^{(2)}\left(\frac{\mu^2}{m^2}, \frac{\mu^2}{s_{ij}} \right) + \mathcal{O}(\as^3) \,,
\end{equation}
where the one-loop and two-loop coefficients are 
\begingroup
\allowdisplaybreaks
\begin{align}
	\boldsymbol{\mathcal{F}}_{[c]}^{(1)} &= \mathcal{F}^{(1)} + \mathcal{F}_{[c]}^{(1)}  \,,\\
	\boldsymbol{\mathcal{F}}_{[c]}^{(2)} &= \mathcal{F}^{(2)} + \SSoft^{(2),\epsilon^0} + \mathcal{F}_{[c]}^{(2)} 
	+ n_h\left( \frac{\pi^2}{18} +\frac{1}{3}l^2_{\mu m} \right) \ZNeubert_{m \ll \mu_h}^{(1), 1/\epsilon} + n_h\left( -\frac{2}{9}\zeta_3 + \frac{\pi^2}{18}l_{\mu m} +\frac{1}{9}l^3_{\mu m}  \right) \ZNeubert_{m \ll \mu_h}^{(1), 1/\epsilon^2} \,.
\end{align}
\endgroup
The functions \mbox{$\mathcal{F}^{(1)}$} and \mbox{$\mathcal{F}^{(2)}$} are process-independent and given by
\begingroup
\allowdisplaybreaks
\begin{align}
	\mathcal{F}^{(1)} &= Z_{[\mathcal{Q}]}^{(1),\epsilon^0} \,,\\
	\mathcal{F}^{(2)} &= Z_{[\mathcal{Q}]}^{(2),\epsilon^0} - Z_{[\mathcal{Q}]}^{(1),1/\epsilon} Z_{[\mathcal{Q}]}^{(1),\epsilon} - Z_{[\mathcal{Q}]}^{(1),1/\epsilon^2} Z_{[\mathcal{Q}]}^{(1),\epsilon^2}  \,,
	\label{master_equation6}
\end{align}
\endgroup
where \mbox{$Z_{[\mathcal{\mathcal{Q}}]}^{(k),\epsilon^j}$} (\mbox{$k = 1,2$} and \mbox{$j \in \{-2,-1,0,1,2\}$}) depends on the ratio \mbox{$\frac{\mu^2}{m^2}$} and represents the coefficient of the $\epsilon^j$ term in the Laurent expansion of \mbox{$Z_{[Q]}^{(k)}$}.
The channel-dependent contributions \mbox{$\mathcal{F}_{[c]}^{(1)}$} and \mbox{$\mathcal{F}_{[c]}^{(2)}$} are
\begingroup
\allowdisplaybreaks
\begin{align}
	\mathcal{F}_{[q]}^{(1)} &= 0 \,,~~~~~~~~~~~~ \mathcal{F}_{[g]}^{(1)} = Z_{[g]}^{(1),\epsilon^0}  \,,\\
	\mathcal{F}_{[q]}^{(2)} &= Z_{[q]}^{(2),\epsilon^0} \,,~~~~~ \mathcal{F}_{[g]}^{(2)} =  Z_{[g]}^{(2),\epsilon^0} - Z_{[g]}^{(1),1/\epsilon} Z_{[g]}^{(1),\epsilon} + Z_{[g]}^{(1),\epsilon^0} Z_{[\mathcal{Q}]}^{(1),\epsilon^0} \,,
\end{align}
\endgroup
while the additional $n_h$-dependent terms proportional to the pole coefficients of the one-loop operator \mbox{$\ZNeubert_{m \ll \mu_h}^{(1)}$} are generated by the application of the inverse decoupling\footnote{The inverse decoupling relation at one-loop order is obtained by inverting Eq.~\eqref{oneloop_decoupling}.}  \mbox{$\alphasNl \to \alphasNf$} to the expression of the subtraction operator $\ZNeubert$ of Ref.~\cite{Ferroglia:2009ii}. 
In this reference, $\ZNeubert$ is given in the effective theory with $n_h$ decoupled quarks. 
However, we observe that, in the full theory with \mbox{$n_f = n_l + n_h$} flavours, such an operator contains not only $\epsilon$ poles but also finite terms starting from two-loop order.

The soft function $\SSoft$ in Eq.~\eqref{master_equation4} corresponds to that computed in Ref.~\cite{Wang:2023qbf}. 
For convenience, we rewrite Eq.~\eqref{two-loop_soft-function} as
\begin{equation}
	\SSoft\left(\alphasNf,\frac{\mu^2}{s_{ij}},\frac{\mu^2}{m^2} , \epsilon\right) = 1 + \left(\frac{\alphasNf(\mu^2)}{4\pi}\right)^{\!2} \!\SSoft^{(2)}\left(\frac{\mu^2}{s_{ij}},\frac{\mu^2}{m^2} , \epsilon\right) + \mathcal{O}(\as^3) \,,
\end{equation}
where the two-loop colour operator $\SSoft^{(2)}$, for the process in Eq.~\eqref{HQQ_process}, reads
\begin{align}
	\SSoft^{(2)}\left(\frac{\mu^2}{s_{ij}},\frac{\mu^2}{m^2} , \epsilon\right) &= 
	\biggl[ (\mathbf{T}_1^2 + C_F)l_{\mu m}
	+  \sum_{i < j} \mathbf{T}_i \cdot \mathbf{T}_j \ln \left( \frac{\mu^2}{-s_{ij}} \right)
	\biggr] \notag \\
	&\times  T_R n_h \left( -\frac{4}{3 \epsilon^2} + \frac{1}{\epsilon}\left( \frac{20}{9} -\frac{8}{3}l_{\mu m} \right) - \frac{8}{3}l^2_{\mu m} + \frac{40}{9}l_{\mu m}  -\frac{112}{27} - \frac{4}{3} \zeta_2  + \mathcal{O}(\epsilon) \right) \,,
\end{align}
with \mbox{$\mathbf{T}_1 + \mathbf{T}_2 + \mathbf{T}_3 + \mathbf{T}_4 = 0$} due to colour conservation.
The invariants $s_{ij}$, defined in Eq.~\eqref{eq:sij_inv}, are evaluated on the set of massless momenta \mbox{$\{\tilde{p}_i \}_{i=1,\ldots,4}$} obtained by projecting the corresponding massive momenta \mbox{$\{p_i\}_{i=1,\ldots,4}$}.
Further details on the mappings are provided in Appendix \ref{app:mapping}.

By expanding both sides of Eq.~\eqref{master_equation4} in $\alphasNf/(4\pi)$, we obtain the one-loop and two-loop relations between the massive and massless finite remainders
\begingroup
\allowdisplaybreaks
\begin{align}
	\ket{ \M_{m}^{(1),\finSCET} } &= \ket{ \M_{(m=0)}^{(1),\finSCET} } + \boldsymbol{\mathcal{F}}^{(1)}_{[c]} \ket{ \M_{(m=0)}^{(0)} } \,, \\
	\ket{ \M_{m}^{(2),\finSCET} } &= \ket{ \M_{(m=0)}^{(2),\finSCET} } + \boldsymbol{\mathcal{F}}^{(1)}_{[c]}  \ket{ \M_{(m=0)}^{(1),\finSCET} } + \boldsymbol{\mathcal{F}}^{(2)}_{[c]}  \ket{ \M_{(m=0)}^{(0)} } \,.
	\label{master_equation3}
\end{align}
\endgroup
As a final step, to be consistent with an NNLO computation performed in an $n_l$-flavour scheme, we need to decouple $n_h$ heavy quarks from the running of $\as$. This is achieved by applying to Eq.~\eqref{master_equation5} the finite shift given in Eq.~\eqref{eq:twoloop_decoupling} (see Appendix \ref{app:soft}), which is responsible for introducing additional logarithmic terms in the heavy-quark mass $m$. 

To summarise, we can construct an approximate version of the massive two-loop finite remainder \mbox{$\ket{ \M_{m}^{(2),\finSCET} }$} starting from the corresponding massless one \mbox{$\ket{ \M_{(m=0)}^{(2),\finSCET} }$}, without relying on the higher orders in $\epsilon$ of the massless one-loop UV renormalised amplitudes.
A similar approach was adopted in Ref.~\cite{Mazzitelli:2024ura}.

The procedure outlined in this section has been implemented in a private {\scshape C++} library called \HQQAmp.
The library takes as input the set of massless momenta $\{\tilde{p}_i\}$ and the renormalisation scale $\mu$, and it returns the massive one-loop or two-loop finite remainder for a given partonic channel $c\bar c$ in Eq.~\eqref{HQQ_process}.
In our implementation, \mbox{$\ket{ \M_{(m=0)}^{(2),\finSCET} }$} is evaluated in leading-colour approximation, starting from the ancillary files provided by the authors of Ref.~\cite{Badger:2021ega}. In contrast, all other terms in Eq.~\eqref{master_equation3}, which contain logarithmically enhanced contributions in $m$, are evaluated in full colour.
Indeed, the library relies on \OpenLoops~\cite{Cascioli:2011va, Buccioni:2017yxi,Buccioni:2019sur} for the evaluation of the full-colour tree-level and one-loop matrix elements and on the \Pentagon library~\cite{Chicherin:2021dyp} for the evaluation of the special functions into which the MIs present in the leading-colour two-loop amplitudes of Ref.~\cite{Badger:2021ega} are decomposed.

\section{Validation}
\label{sec:validation}

In this section, we study the quality of our approximations for the virtual contribution at NLO and extend these results to NNLO. 
Finally, we present a combination of the two double-virtual approximations and an estimate of the associated uncertainty.

Our computation of the NNLO cross section is implemented within the \Matrix framework~\cite{Grazzini:2017mhc}, where IR singularities are handled and cancelled via a process-independent implementation of the $q_T$-subtraction formalism~\cite{Catani:2007vq}, extended to heavy-quark production in Refs.~\cite{Bonciani:2015sha,Catani:2019iny,Catani:2019hip,Catani:2023tby}.
All NLO-like singularities are treated by dipole subtraction~\cite{Catani:1996jh,Catani:1996vz,Catani:2002hc,Kallweit:2017khh,Dittmaier:1999mb,Dittmaier:2008md,Gehrmann:2010ry,Schonherr:2017qcj}.

The required tree-level and one-loop amplitudes are obtained via
\OpenLoops~\cite{Cascioli:2011va, Buccioni:2017yxi,Buccioni:2019sur}
and \Recola~\cite{Actis:2012qn,Actis:2016mpe,Denner:2017wsf}.
Operatively, a technical cut-off \mbox{$\rcut \equiv q_T^{\mathrm{cut}}/Q$} is introduced on the dimensionless
variable \mbox{$q_T/Q$}, where $q_T (Q)$ is the transverse momentum (invariant mass) of the \ttH system.
The final result, which corresponds to the limit \mbox{$\rcut\to 0$}, is extracted by simultaneously computing
the cross section at fixed values of $\rcut$ and then performing an \mbox{$\rcut\to 0$}
extrapolation. 
More details on the procedure and its uncertainties can be found in
Refs.~\cite{Grazzini:2017mhc,Catani:2021cbl}.

We consider proton--proton collisions at the centre-of-mass energy \mbox{$\sqrt{s}=13.6\,\mathrm{TeV}$}.
We use the NNLO NNPDF40~\cite{NNPDF:2021njg} parton distribution functions~(PDFs)
throughout, with the photon PDF included \cite{NNPDF:2024djq}, according to the LUXqed methodology \cite{Manohar:2016nzj}.\footnote{The specific PDF set we use is {\tt NNPDF40\_nnlo\_as\_01180\_qed}.} The QCD running coupling $\as$ is consistently evaluated at three-loop order. 

The pole mass of the top quark is \mbox{$m_t=172.5$\,GeV}, while the Higgs boson mass is \mbox{$m_H=125.09$\,GeV}.
We use the $G_{\mu}$ scheme with \mbox{$G_F = 1.16639 \cdot 10^{-5} \,\mathrm{GeV}^{-2} $} and the following values for the gauge-boson masses and widths:
\begin{align}
	m_W &= 80.379 \,\mathrm{GeV}\,,~~~~~~\Gamma_W = 2.085 \,\mathrm{GeV}\,,\nonumber\\
	m_Z &= 91.1876 \,\mathrm{GeV}\,,~~~~~\Gamma_Z = 2.4952 \,\mathrm{GeV}\, .
\end{align}
We use the complex-mass scheme \cite{Denner:2005fg,Buccioni:2019sur} in which width effects are absorbed into the complex-valued renormalised squared masses
\begin{equation}
  \mu_j^2=m_j^2-i\Gamma_j m_j~\,,~~~~~~~j = W,Z\, .
\end{equation}
The EW coupling is derived from the gauge-boson masses using
\begin{equation}
  	\alpha = \frac{\sqrt 2}{\pi} G_F \left|\mu_W^2 \left( 1-\sin^2\theta_W \right)\right| \,,
\end{equation}
and the complex-valued weak mixing angle is given by
\begin{equation}
\sin^2\theta_W=1-\frac{\mu_W^2}{\mu_Z^2}\, .
\end{equation}
We consider a diagonal CKM matrix.

Unless stated otherwise, the central values of the renormalisation and factorisation scales, $\muR$ and $\muF$, are fixed at
\begin{equation}
	\muR=\muF=(E_{T,t} + E_{T, \bar t} + E_{T, H})/2 \,,
\end{equation}
where \mbox{$E_{T,i} = \sqrt{ m_i^2 + p_{T,i}^2}$} is the transverse energy of a massive particle, \mbox{$i \in \{t,\bar t, H\}$}. 
This dynamic scale reduces to the fixed scale \mbox{$(m_t +m_H/2)$} in the limit \mbox{$p_T \to 0$} for the top quarks and the Higgs boson. 

The two-loop virtual amplitude contributes to the NNLO calculation through its interference with the Born amplitude. 
To isolate the part that requires an approximation, 
we define the hard-virtual coefficient
\begin{equation}
\label{eq:H}
	H^{(n)}(\muIR)=\left.\frac{2{\mathrm{Re}}\left(\M^{(n), \mathrm{fin}}(\muIR,\muR)\M^{(0)*}\right)}{|\M^{(0)}|^2}\right\vert_{\muR=Q}\,, ~~~~~~~~~~n=1,2,\dots,
\end{equation}
which is computed through the interference of the Born amplitude ${\cal M}^{(0)}$ for the \mbox{$c{\bar c}\to t{\bar t}H$} process (\mbox{$c=q,g$}) with the $n$-loop IR subtracted virtual correction \mbox{$\M^{(n), \mathrm{fin}}(\muIR,\muR)$} (in an expansion in powers of $\as/(2 \pi)$), evaluated within the scheme of Ref.~\cite{Ferroglia:2009ii} at the subtraction scale $\muIR$.
In our framework, the contribution of the $H^{(n)}$ coefficient to the N$^{(n)}$LO cross section reads
\begin{equation}
  d\sigma_{H^{(n)}} \equiv \left(\frac{\as(\mu_R)}{2\pi}\right)^{\!n} H^{(n)}(Q)\, d\sigma_{\rm LO} \,,
\end{equation}
where a summation over the $q{\bar q}$ and $gg$ partonic channels is understood.
Unless stated otherwise, the subtraction scale $\muIR$ is always set to the invariant mass $Q$ of the event.
We remind the reader that this term represents the only approximated contribution to the NNLO cross section. 
All the remaining ingredients, including the one-loop squared contribution to the diagonal channels, are evaluated exactly. 

We define our soft approximation (SA) of the $H^{(n)}$ coefficient as
\begin{equation}
\label{eq:Hn_SA}
	H^{(n)}_{\SA}(\muIR)=\left.\frac{2{\mathrm{Re}}\left(\M^{(n), \mathrm{fin}}_{\SA}(\muIR,\muR)\M^{(0)*}_{\SA}\right)}{|\M^{(0)}_{\SA}|^2}\right\vert_{\muR={\widetilde Q}}\,, ~~~~~~~~~~n=1,2,\dots,
\end{equation}
where ${\widetilde Q}$ is the virtuality of the $t{\bar t}$ pair.
The finite remainder \mbox{$\M^{(n), \mathrm{fin}}_{\SA}$ }and the Born amplitude \mbox{$\M^{(0)}_{\SA}$} in the soft approximation are obtained by applying the factorisation formula in Eq.~\eqref{eq:fact}. 
To achieve this, we need to introduce a prescription that defines how to transform an event containing a $t \bar t$ pair and a Higgs boson into a corresponding event where the Higgs boson is removed.
As in Ref.~\cite{Catani:2022mfv}, we exploit the $q_T$-recoil prescription \cite{Catani:2015vma} where the momenta of the top and anti-top
quarks are left unchanged and the initial-state partons equally reabsorb the transverse momentum
of the Higgs boson. 
This prescription allows us to evaluate the (non-radiative) $t \bar t$ amplitudes on the r.h.s.\ of the factorisation formula in Eq.~\eqref{eq:fact}.

Analogously to Eq.~\eqref{eq:Hn_SA}, we define the hard-virtual coefficient in the massification approach (MA) as
\begin{equation}
\label{eq:Hn_MA}
	H^{(n)}_{\MA}(\muIR)=\left.\frac{2{\mathrm{Re}}\left(\M^{(n), \mathrm{fin}}_{\MA}(\muIR,\muR)\M^{(0)*}_{\MA}\right)}{|\M^{(0)}_{\MA}|^2}\right\vert_{\muR={\widetilde Q}}\,, ~~~~~~~~~~n=1,2,\dots,
\end{equation}
where the massive finite remainder \mbox{$\M^{(n), \mathrm{fin}}_{\MA}$} is obtained by applying Eq.~\eqref{master_equation5} to the massless finite remainder of Ref.~\cite{Badger:2021ega}.
It is worth mentioning that, contrary to Ref.~\cite{Buonocore:2023ljm}, we include contributions due to massive top-quark loops as described in Sec.~\ref{subsubsec:massification_HQQ}. However, their impact on the two-loop hard-virtual coefficient is found to be subleading compared to the systematic error we will assign to our approximation. 
To evaluate the massless finite remainders appearing on the r.h.s.\ of Eq.~\eqref{master_equation5} and to ensure momentum conservation, we need a prescription that projects a \ttH event onto an event with massless top quarks.
Since the $q\bar q$ and $gg$ partonic channels have distinct leading-order momentum flows, we use a dedicated mapping for each channel. 
In particular, the choice of the mapping is quite delicate for the $gg$ channel, as an initial-state collinear singularity (absent in the massive case) can emerge in the massless limit of the amplitude when the (anti)top-quark transverse momentum vanishes. 
Although our mapping is designed to screen this divergence, we anticipate that sensitivity to this (unphysical) effect will still lead to large uncertainties at small $p_T$. For more details on the choice of the mappings, we refer the reader to Appendix~\ref{app:mapping}.
In Eq.~\eqref{eq:Hn_MA}, $\widetilde Q$ denotes the invariant mass of the projected event. 

We point out that our approximations in Eqs.~\eqref{eq:Hn_SA} and \eqref{eq:Hn_MA} are carried out on {\it both} the numerator and the denominator in the definition of $H^{(n)}$. 
Effectively, this \textit{reweighting} procedure corresponds to a rescaling of the approximated $n$-loop finite remainder by the exact Born amplitude, thus significantly improving the quality of the approximations. 
A similar improvement is observed in other contexts, such as the large-$m_t$ approximation in Higgs boson production via gluon fusion, both at the inclusive~\cite{Kramer:1996iq} and at the differential~\cite{Jones:2018hbb} level.
\begin{figure*}[t]
  \includegraphics[height=0.38\textheight]{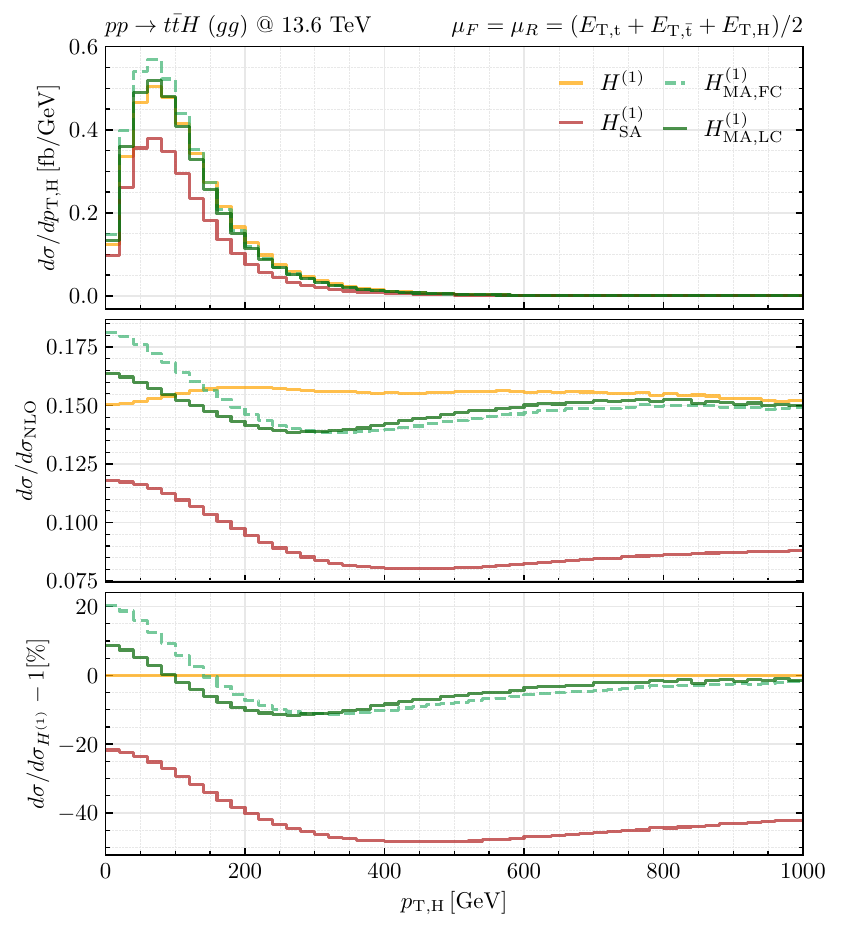}\hfill
  \includegraphics[height=0.38\textheight]{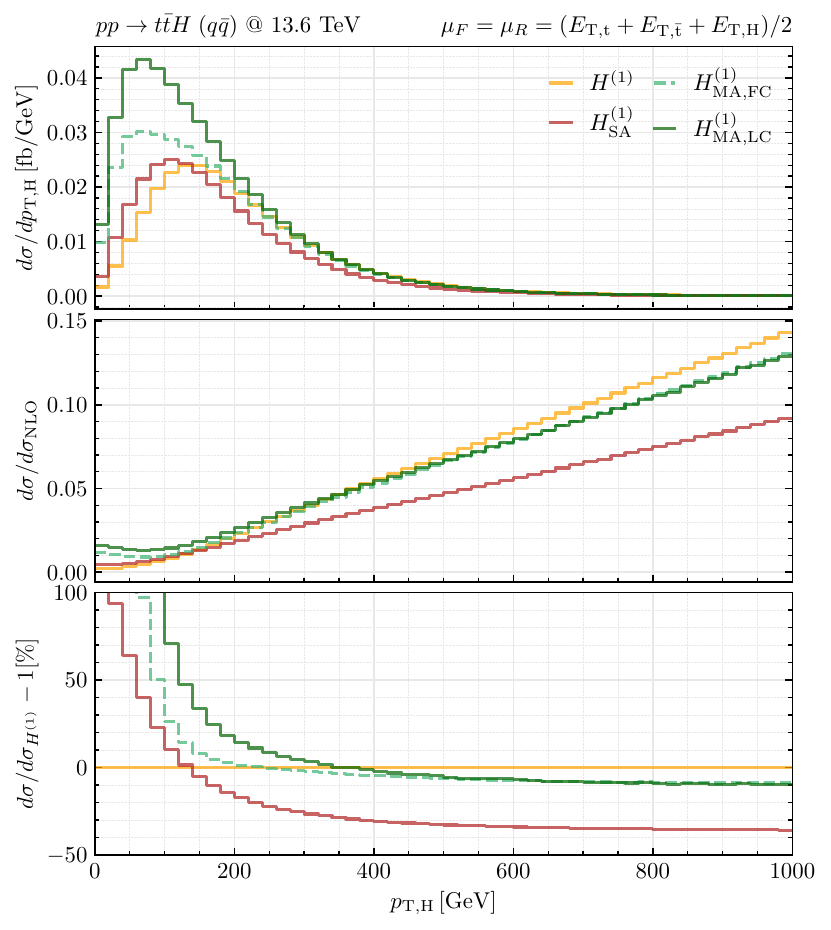}
  \caption{Hard-virtual contribution to the NLO correction, $\sigma_{\hspace{-0.05cm}H^{(1)}}$, as a function of the transverse momentum of the Higgs boson. 
  For the $gg$ (left) and $q \bar q$ (right) partonic channels, we display the comparison between the exact result (yellow curve) and those obtained by applying two approximations: a soft Higgs boson approximation (red curve) and the high-energy expansion (green curves). Two implementations of the latter are considered: a full-colour construction, dubbed FC (dashed green curve), and a construction, labelled LC, in which the one-loop massless finite remainder is evaluated in leading-colour approximation, while the logarithmically enhanced terms in the top-quark mass are kept in full colour (solid dark-green curve).
  The central panels display the impact of the hard-virtual contribution on the NLO cross section, while the lower panels illustrate the quality of the one-loop approximation.}
    \label{fig:tth_H1_SA-MA}
\end{figure*}

We begin our analysis by examining the hard-virtual contribution to the differential cross section in the transverse momentum $p_{\rm T,H}$ of the Higgs boson.
In Fig.~\ref{fig:tth_H1_SA-MA}, we present the one-loop results for the $gg$ (left) and $q{\bar q}$ (right) partonic channels. 
The upper panels show the comparison between the exact $\sigma_{\hspace{-0.05cm}H^{(1)}}$ contribution and the corresponding results obtained by applying the two approximations mentioned above. 
Two implementations of the massification approach are considered: a full-colour construction (dashed green curves), labelled FC, and a construction, labelled LC, in which the one-loop massless finite remainders are evaluated in leading-colour approximation, while the logarithmically enhanced terms in the top-quark mass are kept in full colour (solid dark-green curves). This second version aims to mirror the construction of the massive two-loop finite remainders, which has to rely on massless two-loop finite remainders that are not available beyond leading colour.
As mentioned in Sec.~\ref{subsubsec:massification_HQQ}, the logarithmically enhanced terms in the mass are always evaluated in full colour to ensure their exact cancellation with analogous logarithms arising from the real contribution.
The central panels display the one-loop results normalised to the whole NLO cross section, while the lower panels show the ratio of the approximated results to the exact hard-virtual contribution.

From the central panels, we observe that in the $gg$ channel the impact of the hard-virtual contribution (yellow curve) is relatively uniform across the $p_{\rm T,H}$ spectrum, contributing approximately $15\%$ to the NLO cross section. In contrast, for the $q{\bar q}$ channel, the hard-virtual contribution is negligible at low $p_{\rm T,H}$, remaining at the percent level in the intermediate region, but increases to around $14\%$ at \mbox{$p_{\rm T,H}=1~{\rm TeV}$}.
The lower panels show the quality of the approximation of the hard-virtual contribution. 
In the $gg$ channel, we observe that the MA results reproduce the exact one-loop contribution within $\pm (10-20)\%$ over the whole range of transverse momenta. 
Differences of $\mathcal{O}(10\%)$ between the two versions of the massification are visible in the low $p_{\rm T,H}$ bins, while they become negligible in the tail where the MA results asymptotically tend to the exact hard-virtual contribution, as expected.
On the contrary, the soft approximation undershoots the correct result from about $25\%$ in the peak region to $40-50\%$ at high $p_{\rm T,H}$. 
In the $q{\bar q}$ channel, the quality of the two approximations is significantly different in the low and high $p_{\rm T,H}$ regions. For \mbox{$p_{\rm T,H}\gtap 200\,{\rm GeV}$}, where the impact of the hard-virtual contribution increases monotonically, the MA result reproduces the exact one-loop contribution within an accuracy of $\pm 10\%$. In contrast, the soft approximation systematically underestimates the exact result by roughly $30-40\%$. In the region \mbox{$p_{\rm T,H}\ltap 200$ GeV}, where the impact of the hard-virtual contribution on the NLO cross section is at the few-percent level, the quality of the two approximations is visibly worse. Discrepancies of $\mathcal{O}(100\%)$ or even larger can be observed, especially in the MA case. In this region, we also notice larger differences between the results obtained with the two MA versions, with FC performing significantly better.

The previous discussion underscores that the two distinct and complementary approximations of the hard-virtual coefficient generally provide a reliable estimate of the virtual contribution across a wide range of the $p_{\rm T,H}$ spectrum.
A similar conclusion can be drawn from the study of other differential distributions.
Nevertheless, it is clear that the quality of these approximations ultimately depends on the relative impact of the hard-virtual contribution on the cross section.

At NNLO, it becomes crucial to devise a method for quantifying the uncertainties associated with each approximation and, subsequently, to establish a robust procedure for combining them.
To this end, our approach builds upon the strategy proposed in Ref.~\cite{Buonocore:2023ljm} for \ttW production.
The starting point is the observation that, regardless of the chosen approximation, the associated uncertainty cannot be expected to be smaller than the relative difference in the approximate and exact hard-virtual coefficients observed at NLO. 
Another critical point is that the approximated result depends on the subtraction scale $\muIR$ at which the approximation is applied. If the exact result were available, the corresponding scale-dependent terms would exactly compensate for any change in the choice of $\muIR$. However, with an approximation, this is no longer the case. The observed variation with respect to $\muIR$ provides an estimate of the resulting uncertainties.
Therefore, for each partonic channel, we define a bin-wise error on $d\sigma_{\hspace{-0.05cm}H^{(2)}}$ separately for the soft approximation and the massification approach, by taking into account 
\begin{enumerate}
	\item the discrepancy between the exact and approximated predictions at NLO;
	\item the effects due to the variation of the subtraction scale $\muIR$, at which the approximations in Eqs.~(\ref{eq:Hn_SA}) and (\ref{eq:Hn_MA}) are applied, by a factor of two around the central scale $Q$ (adding the exact two-loop evolution from $\muIR$ to Q). 
\end{enumerate}
The first source of error, dubbed \textit{$H^{(1)}$-based error}, results from the assumption that, as discussed above, the relative uncertainty assigned to each approximation is not smaller than the relative deviation of the respective approximate result for $\sigma_{\hspace{-0.03cm}H^{(1)}}$ from the exact.
Therefore, we define the $H^{(1)}$-based errors, $\delta_{\SA}^{H^{(1)}}$ and $\delta_{\MA}^{H^{(1)}}$, as
\begingroup
\allowdisplaybreaks
\begin{subequations}
\label{eq:H1-based_errors}
\begin{align}
	\delta_{\SA}^{H^{(1)}} &= 2 \times \left| \frac{\sigma_{\hspace{-0.05cm}H^{(1)}_{\SA}}}{\sigma_{\hspace{-0.05cm}H^{(1)}}} -1 \right| \times \max\left( | \sigma_{\hspace{-0.05cm}H^{(2)}_{\SA}} |, | \sigma_{\hspace{-0.05cm}H^{(2)}_{\MA}}| \right) \,, \label{eq:H1-based_errors_a} \\
	\delta_{\MA}^{H^{(1)}} &= 2 \times \max\left( \left| \frac{\sigma_{\hspace{-0.05cm}H^{(1)}_{\mathrm{MA, FC}}}}{\sigma_{\hspace{-0.05cm}H^{(1)}}} -1 \right|, \left| \frac{\sigma_{\hspace{-0.05cm}H^{(1)}_{\mathrm{MA, LC}}}}{\sigma_{\hspace{-0.05cm}H^{(1)}}} -1 \right| \right) \times \max\left( | \sigma_{\hspace{-0.05cm}H^{(2)}_{\SA}}|, | \sigma_{\hspace{-0.05cm}H^{(2)}_{\MA}}| \right) \label{eq:H1-based_errors_b} \,. 
\end{align}
\end{subequations}
\endgroup
In Eq.~(\ref{eq:H1-based_errors_a}) the relative deviation for SA, \mbox{$\left| \sigma_{\hspace{-0.05cm}H^{(1)}_{\SA}}/\sigma_{\hspace{-0.05cm}H^{(1)}} -1 \right|$} (or analogously the maximum of the two deviations in FC and LC in Eq.~(\ref{eq:H1-based_errors_b}) for MA) is multiplied by the maximum of $\sigma_{\hspace{-0.05cm}H^{(2)}_{\SA}}$ and $\sigma_{\hspace{-0.05cm}H^{(2)}_{\MA}}$.
Such \textit{correlated} procedure, which we apply to our differential predictions,
prevents us from underrepresenting the $H^{(1)}$-based error in phase space regions where the double-virtual contribution in one of the two approximations is seriously underestimated.\footnote{For simplicity, the procedure is kept \textit{uncorrelated} only for total cross sections.} The result is that, by construction, the $H^{(1)}$-based error assigns a larger systematic uncertainty to the approximation of the double-virtual contribution where $\sigma_{\hspace{-0.05cm}H^{(1)}}$ agrees worse. This happens, for example, for the soft approximation in the high-$p_T$ region.  The final error is conservatively multiplied by a factor of two in both cases.

The second source of error, labelled \textit{$\muIR$-variation error}, is defined as
\begingroup
\allowdisplaybreaks
\begin{subequations}
\label{eq:muIR-variation_errors}
\begin{align}
	\hspace{-0.5cm} \delta_{\SA}^{\mu_{\mathrm{IR}}} &= \max\left( \left| \sigma_{\hspace{-0.05cm}H^{(2)}_{\SA}(\widetilde{Q}/2)}+ (Q/2 \to Q) - \sigma_{\hspace{-0.05cm}H^{(2)}_{\SA}} \right|, \left| \sigma_{\hspace{-0.05cm}H^{(2)}_{\SA}(2\widetilde{Q})} + (2Q \to Q) - \sigma_{\hspace{-0.05cm}H^{(2)}_{\SA}} \right| \right) \,, \label{eq:SA_muIR-variation_errors} \\
	\hspace{-0.5cm} \delta_{\MA}^{\mu_{\mathrm{IR}}}  &= \max\left( \left| \sigma_{\hspace{-0.05cm}H^{(2)}_{\MA}(\widetilde{Q}/2)} + (Q/2 \to Q) - \sigma_{\hspace{-0.05cm}H^{(2)}_{\MA}}\right|, \left| \sigma_{\hspace{-0.05cm}H^{(2)}_{\MA}(2\widetilde{Q})} + (2Q \to Q) - \sigma_{\hspace{-0.05cm}H^{(2)}_{\MA}} \right| \right) \,, \label{eq:MA_muIR-variation_errors}
\end{align}
\end{subequations}
\endgroup
where we use the abbreviation \mbox{$(\muIR \to Q)$} to collectively indicate the terms from the running of the two-loop amplitudes based on the exact tree-level and one-loop $\ttH$ amplitudes. 
This error estimate takes into account the arbitrariness in the choice of the scale $\muIR$ at which the approximation is applied. 

\begin{figure*}[t]
  \includegraphics[height=0.38\textheight]{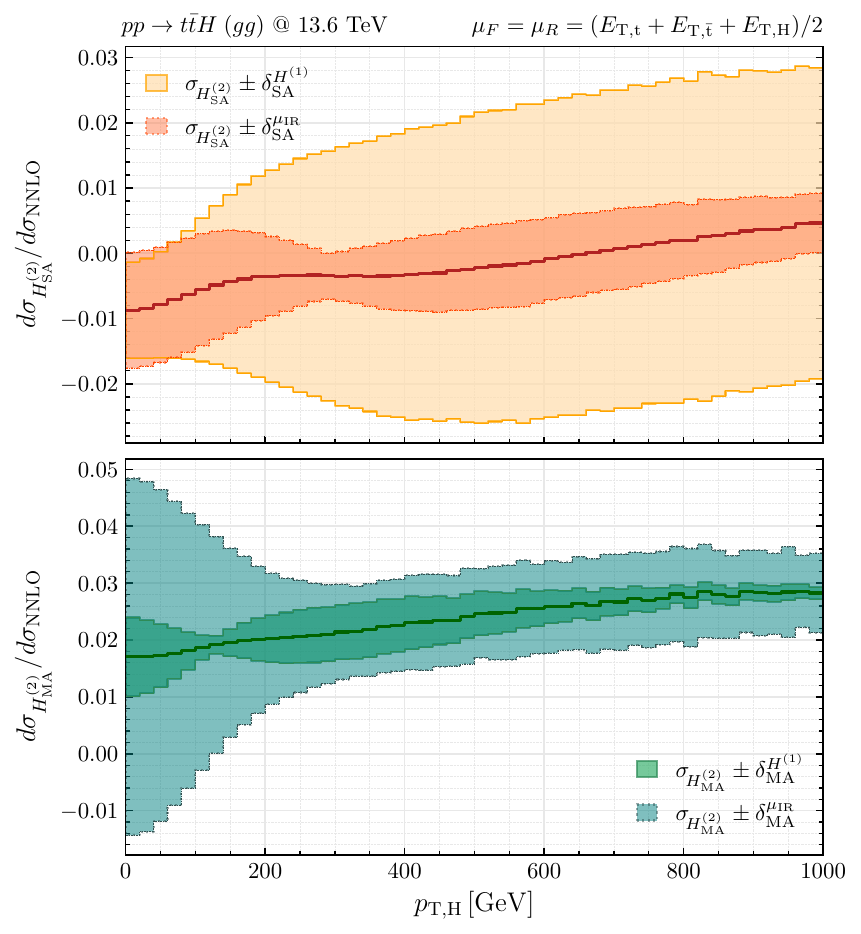}\hfill
  \includegraphics[height=0.38\textheight]{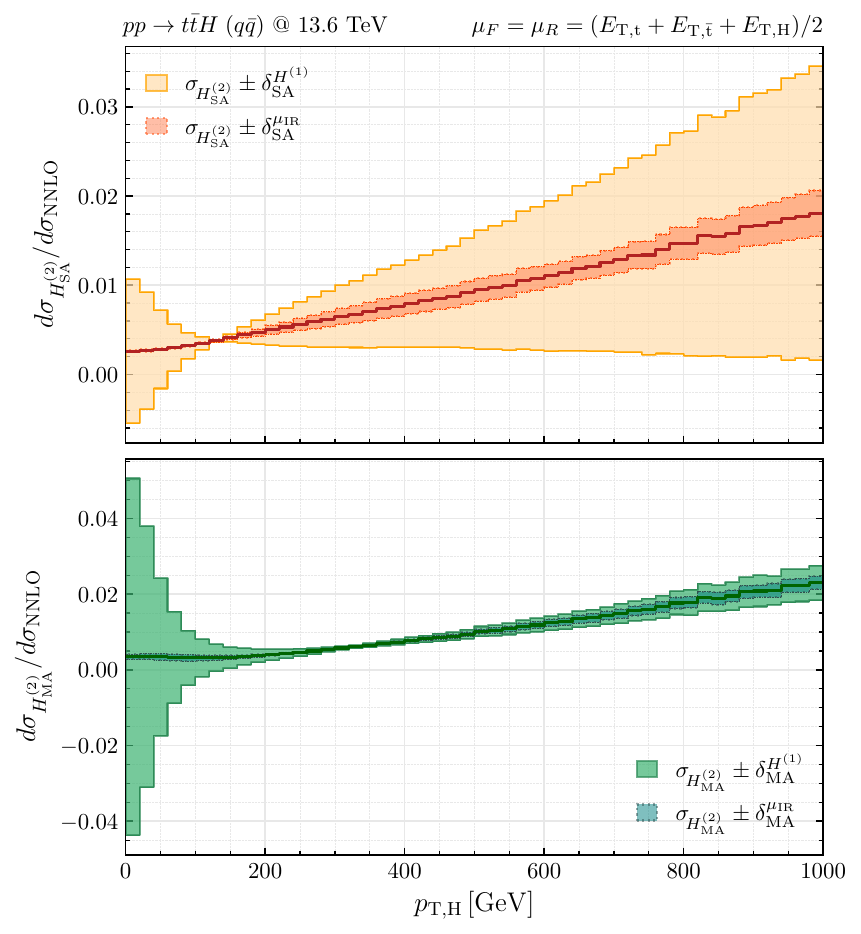}
  \caption{Study of the different sources (i.e.\ \textit{$H^{(1)}$-based} and \textit{$\muIR$-variation} error) that contribute to the systematic uncertainty we assign to the hard-virtual contribution at NNLO, $\sigma_{\hspace{-0.05cm}H^{(2)}}$. 
We separately display $gg$ (left) and $q \bar q$ (right) partonic channels. For each partonic channel, we show the errors assigned to the soft (upper panel) and high-energy (lower panel) approximations of the two-loop hard-virtual contribution before their combination (see main text for further details).}
    \label{fig:tth_H2_SA-MA}
\end{figure*}
%
In Fig.~\ref{fig:tth_H2_SA-MA} we show the relative impact of the approximated $\sigma_{\hspace{-0.05cm}H^{(2)}}$ contributions with the two sources of uncertainties for the $gg$ (left) and $q{\bar q}$ (right) channels.
In the $gg$ channel, for the soft approximation (upper panel), the $\mu_{\mathrm{IR}}$-uncertainty dominates for \mbox{$p_{\rm T,H}\ltap 70\,{\rm GeV}$}, while at large $p_{\rm T,H}$ the $H^{(1)}$-based uncertainty is significantly larger. 
For the MA result (lower panel), the $\mu_{\mathrm{IR}}$-variation error dominates over the whole range and, as anticipated previously, is particularly large in the low-$p_{\rm T,H}$ region. 
For the $q{\bar q}$ channel, we find that the $H^{(1)}$-based uncertainty consistently exceeds the $\mu_{\mathrm{IR}}$-uncertainty, particularly at high $p_{\rm T,H}$ in the soft approximation and at low $p_{\rm T,H}$ in the MA result. The plots in Fig.~\ref{fig:tth_H2_SA-MA} also show that the relative impact of $\sigma_{\hspace{-0.05cm}H^{(2)}}$ in both the $gg$ and $q{\bar q}$ channels remains minor, generally at the percent level. Consequently, even substantial relative uncertainties associated with the approximated double-virtual contribution will have a limited effect on the ultimate NNLO result.

The final systematic uncertainty $\delta$ on each approximation, for each partonic channel, is conservatively defined by taking the maximum of the $H^{(1)}$-based~\eqref{eq:H1-based_errors} and $\muIR$-variation~\eqref{eq:muIR-variation_errors} errors.
\begin{figure*}[t]
  \includegraphics[height=0.38\textheight]{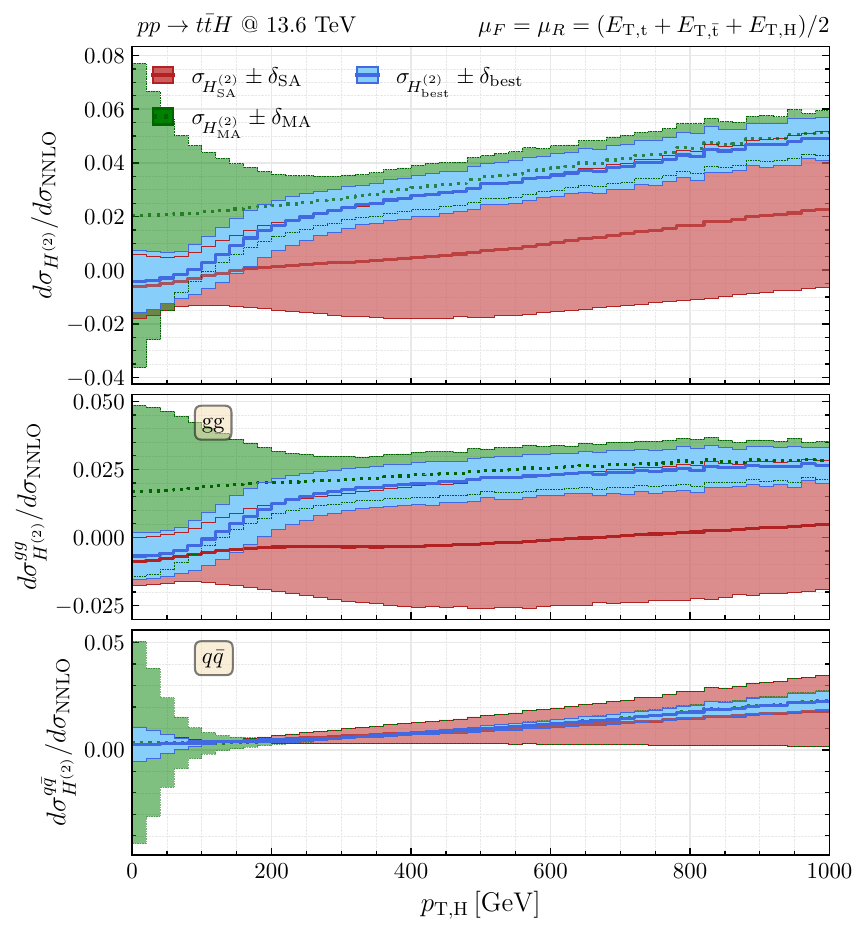}\hfill
  \includegraphics[height=0.38\textheight]{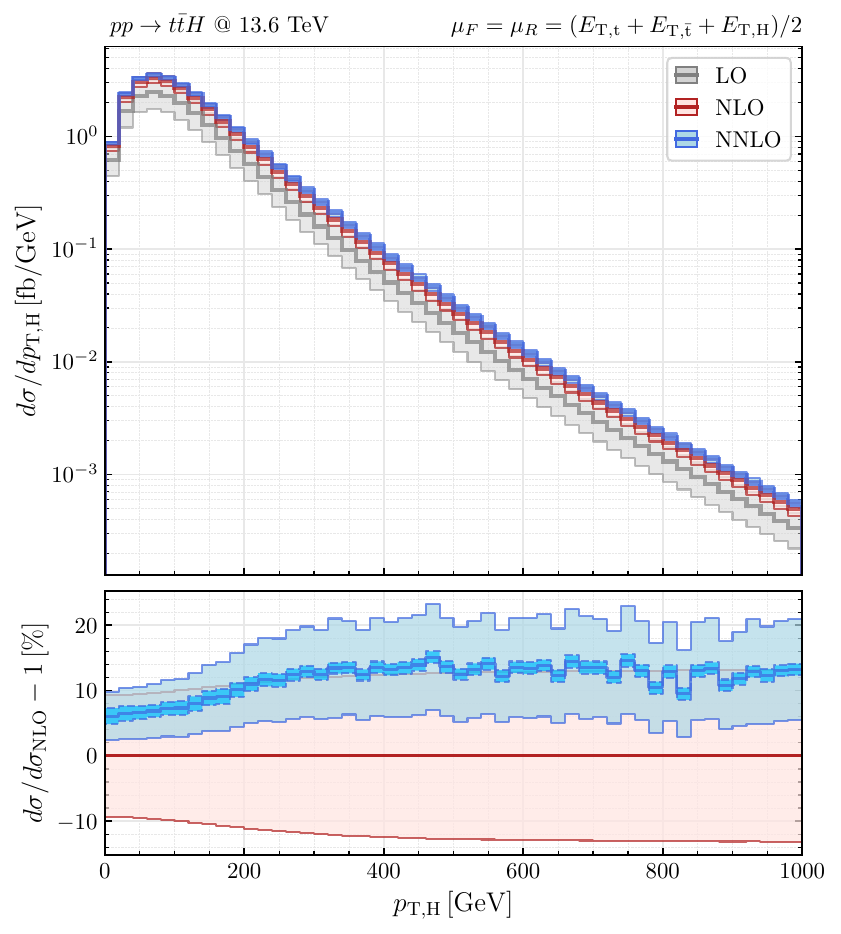}
  \caption{On the left we show the results for $\sigma_{\hspace{-0.09cm}H^{(2)}_{\SA}}$, $\sigma_{\hspace{-0.09cm}H^{(2)}_{\MA}}$ and our \textit{best} prediction $\sigma_{\hspace{-0.09cm}H^{(2)}_{\mathrm{best}}}$, normalised to the NNLO cross section. The prescription adopted to define $\sigma_{\hspace{-0.09cm}H^{(2)}_{\mathrm{best}}}$, and the corresponding systematic error is explained in the main text. 
  On the right, we display the LO, NLO, and NNLO cross sections and their (symmetrised) perturbative uncertainties. The inner NNLO dashed band denotes the systematic uncertainty from the approximation of the double-virtual contribution $\sigma_{\hspace{-0.09cm}H^{(2)}_{\mathrm{best}}}$, based on $\sigma_{\hspace{-0.09cm}H^{(2)}_{\SA}}$ and $\sigma_{\hspace{-0.09cm}H^{(2)}_{\MA}}$.}
    \label{fig:tth_pTh_best}
\end{figure*}
Since we have established the procedure to assign an error separately to each approximation, we can now define our \textit{best} prediction and its corresponding uncertainty in both the $gg$ and $q{\bar q}$ channels by taking the weighted average,
\begingroup
\allowdisplaybreaks
\begin{align}
	\sigma_{\hspace{-0.05cm}H^{(2)}_{\mathrm{best}}} &= \frac{1}{\omega_{\SA} + \omega_{\MA}}\left( \omega_{\SA} \sigma_{\hspace{-0.05cm}H^{(2)}_{\SA}} + \omega_{\MA} \sigma_{\hspace{-0.05cm}H^{(2)}_{\MA}} \right)\,, \\
	\delta_{\mathrm{best}} &= \left( \frac{1}{\omega_{\SA} + \omega_{\MA}} \right)^{\!1/2}\,,
\end{align}
\endgroup
where the weights assigned to each approximation are
\begin{equation}
	 \omega_{\SA} = \frac{1}{\delta_{\SA}^2} ~~~~~\text{and}~~~~~ \omega_{\MA} = \frac{1}{\delta_{\MA}^2} \,.
\end{equation}
Finally, the \textit{best} predictions for the two partonic channels are summed up, and their errors are combined in quadrature.

In Fig.~\ref{fig:tth_pTh_best} (left), we show the combination of the two partonic channels (upper panel), the $gg$ (central panel) and $q \bar q$ (lower panel) channels separately.
In each panel, we display the comparison between the contributions of $\sigma_{\hspace{-0.05cm}H^{(2)}_{\SA}}$, $\sigma_{\hspace{-0.05cm}H^{(2)}_{\MA}}$ and our \textit{best} prediction $\sigma_{\hspace{-0.05cm}H^{(2)}_{\mathrm{best}}}$, normalised to the NNLO cross section, as a function of $p_{\rm T,H}$. The error bands correspond to the systematic uncertainties associated with each approximation. 
As expected, our \textit{best} estimate of the double-virtual contribution is driven by the soft approximation at low $p_{\rm T,H}$ and by the high-energy result in the large $p_{\rm T,H}$ region.

In Fig.~\ref{fig:tth_pTh_best} (right) we show our results for the $p_{\rm T,H}$ spectrum at LO, NLO and NNLO with their perturbative uncertainties.
The latter are evaluated as follows. We start from the customary procedure of independently varying the renormalisation ($\muR$) and factorisation ($\muF$) scales by a factor of 2 around their central value with the constraint \mbox{$0.5\leq \muR/\muF\leq 2$}.
Since the uncertainties estimated this way are typically rather asymmetric, to estimate the residual perturbative uncertainties, we follow Ref.~\cite{Catani:2022mfv} and consider their symmetrised version. More precisely, we take the maximum among the upward and downward variations and assign it symmetrically to construct our final uncertainty, leaving the central prediction unchanged.
The lower panel shows the results normalised to the central NLO prediction.
The dark-blue band represents the uncertainty from the missing knowledge of the exact two-loop virtual contribution, estimated as discussed above.
We observe that the final impact of this uncertainty is significantly smaller than the residual perturbative uncertainties.

\section{Results}
Having discussed our approach to approximate the two-loop virtual contribution and the associated uncertainties, we present numerical predictions for the LHC at \mbox{$\sqrt{s}=13.6\,\mathrm{TeV}$}. The NNLO results in this section correspond to the prediction denoted as \textit{best} in Sec.~\ref{sec:validation}.

In addition to pure QCD predictions, which we label as $\mathrm{N}^i\mathrm{LO}_{\mathrm{QCD}}$ with \mbox{$i = 0,1,2$}, we additively combine our newly computed NNLO QCD results with the full tower of EW corrections up to NLO.\footnote{The calculation of the EW corrections has been validated against a recent implementation in \Whizard~\cite{Bredt:2022nkq}.}
The latter, which represents our most accurate prediction, is dubbed \mbox{$\mathrm{NNLO}_{\mathrm{QCD}}+\mathrm{NLO}_{\mathrm{EW}}$}.

\subsection{Total cross section}
\label{sec:totalXS}
\noindent We start by discussing the pattern of perturbative corrections for the total cross section. Our setup is that described in Sec.~\ref{sec:validation}. The only difference lies in the choice of the central renormalisation and factorisation scales, which in this case are fixed to
\begin{equation}
\muR = \muF = m_t+m_H/2 \,.
\end{equation}

\begin{table}[b]
\centering
\renewcommand{\arraystretch}{1.5}
\setlength{\tabcolsep}{0.4em}
\begin{tabular}{lllll}
& & \multicolumn{1}{l}{$\sigma$ [fb]} \\
\hline
${\mathrm{LO}_{\mathrm{QCD}}}$ & & 
$423.9\phantom{(0)}\!^{+30.7\%}_{-21.9\%} \text{(scale)}$
\\
${\mathrm{NLO}_{\mathrm{QCD}}}$ & & 
$528.9\phantom{(0)}\!^{\phantom{0}+5.7\%}_{\phantom{0}-9.0\%} \text{(scale)}$
\\
${\mathrm{NNLO}_{\mathrm{QCD}}}$ & & 
$550.7(5)\!^{\phantom{0}+0.9\%}_{\phantom{0}-3.1\%} \text{(scale)} \,{\scriptstyle\pm 0.9\%} \text{(approx)} $
\\
${\mathrm{NNLO}_{\mathrm{QCD}}+\mathrm{NLO}_{\mathrm{EW}}}$ & & 
$562.3(5)\!^{\phantom{0}+1.1\%}_{\phantom{0}-3.2\%} \text{(scale)} \,{\scriptstyle\pm 0.9\%} \text{(approx)} $
\\
\end{tabular}
\caption{\label{tab:totalXS} Total $t\bar{t}H$ production cross section at the LHC for a collider energy of $13.6\,\mathrm{TeV}$. Results at different orders in the QCD perturbative expansion, including the full set of NLO EW corrections, are presented. The 7-point scale-variation uncertainties and the systematic uncertainties arising from the approximation of the two-loop virtual amplitudes are displayed in percentage. The numbers in parentheses represent the numerical uncertainties from the combination of the Monte Carlo integration error and the $q_T$-subtraction systematic uncertainty. The numerical uncertainties on the LO and NLO predictions are not shown since they are completely negligible.}
\end{table}

In Table~\ref{tab:totalXS} we report the results for the total cross section at the different orders in the QCD perturbative expansion and our best prediction including the NLO EW corrections. Scale uncertainties are computed using the customary 7-point variation. As discussed in Sect.~\ref{sec:validation},
since scale uncertainties turn out to be rather asymmetric, we follow Ref.~\cite{Catani:2022mfv} and estimate the residual perturbative uncertainties at each order by considering their
symmetrised version\footnote{Note that, for the sake of thoroughness, in Table~\ref{tab:totalXS} and Table~\ref{tab:pT_h} we report the non symmetrised scale uncertainties. In all the figures we plot their symmetrised version.}. 

The convergence of the QCD perturbative expansion can be appreciated thanks to the inclusion of the NNLO corrections. While the $\mathrm{NLO}_\mathrm{QCD}$ prediction is about $25\%$ larger than the $\mathrm{LO}_\mathrm{QCD}$ result, the NNLO QCD corrections are significantly milder, increasing the NLO result by about $4\%$.
The higher level of precision is also reflected in the scale uncertainties, which are significantly reduced at $\mathrm{NNLO}_\mathrm{QCD}$. We finally note that the $\mathrm{NNLO}_\mathrm{QCD}$ scale uncertainty band is entirely contained in the $\mathrm{NLO}_\mathrm{QCD}$ prediction, suggesting that the NLO uncertainty estimate is indeed reliable.

The uncertainty stemming from the approximation of the two-loop virtual corrections is $\mathcal{O}(1\%)$ and, therefore, smaller than the residual perturbative uncertainties at NNLO.
We note that even a conservative linear combination with the scale uncertainties would lead to a result that significantly improves the precision of the NLO prediction.

For reference, we compare our $\mathrm{NNLO}_\mathrm{QCD}$ prediction to the results previously presented in Ref.~\cite{Catani:2022mfv}, which were solely based on the soft-Higgs approximation. In the setup used in the present paper, the $\mathrm{NNLO}_\mathrm{QCD}$ prediction computed as in Ref.~\cite{Catani:2022mfv} amounts to $548.7(5) \,^{+0.8\%}_{-3.0\%}\text{(scale)} \,{\scriptstyle\pm 0.6\%} \text{(approx)}\,\mathrm{fb}$.
This number differs from our current prediction by $0.4\%$, while the systematic uncertainty obtained via the refined procedure is a bit larger, i.e.\ $0.9\%$ instead of $0.6\%$.
Overall, the two approaches are in good agreement for the computation of the total cross section.

Including the whole tower of EW corrections increases the $\mathrm{NNLO}_\mathrm{QCD}$ cross section by approximately $2\%$.
While this effect is considerably smaller than the $\mathrm{NLO}_\mathrm{QCD}$ scale variations, this is no longer the case at NNLO due to the significant reduction in perturbative uncertainties.  This underscores the importance of incorporating NLO EW corrections at this level of precision.

\subsection{Differential distributions}
\label{sec:differential_results}

We now focus on differential distributions. For all the results that are shown in this section, the setup is that of Sec.~\ref{sec:validation}, including the setting for the central renormalisation and factorisation scales,
\begin{equation}
	\muR=\muF=(E_{T,t} + E_{T, \bar t} + E_{T, H})/2 \,.
\end{equation}
In all figures, we present results for the differential cross section in the upper panel, at $\mathrm{LO}_\mathrm{QCD}$, $\mathrm{NLO}_\mathrm{QCD}$ and $\mathrm{NNLO}_\mathrm{QCD}$, plus the additive combination of the latter with the whole tower of NLO EW corrections, \mbox{$\mathrm{NNLO}_{\mathrm{QCD}}+\mathrm{NLO}_{\mathrm{EW}}$}. 
The central panel shows the ratio to the $\mathrm{NLO}_\mathrm{QCD}$ result and, for clarity, only the $\mathrm{NLO}_\mathrm{QCD}$ and $\mathrm{NNLO}_\mathrm{QCD}$ bands are displayed since we are interested in assessing the impact of the NNLO QCD corrections.
Finally, in the lower panels we show the $\mathrm{NNLO}_\mathrm{QCD}$ and \mbox{$\mathrm{NNLO}_{\mathrm{QCD}}+\mathrm{NLO}_{\mathrm{EW}}$} results, normalised to the former, to visualise the impact of the EW corrections.
As discussed before, the uncertainty bands correspond to our estimate of the perturbative uncertainties obtained from the customary 7-point scale variation after symmetrisation. The darker band shown for the $\mathrm{NNLO}_\mathrm{QCD}$ result denotes the uncertainty due to the approximation of the two-loop virtual corrections. For clarity, this band is only displayed in the central panel.

\begin{figure*}[t] 
   \includegraphics[height=0.38\textheight]{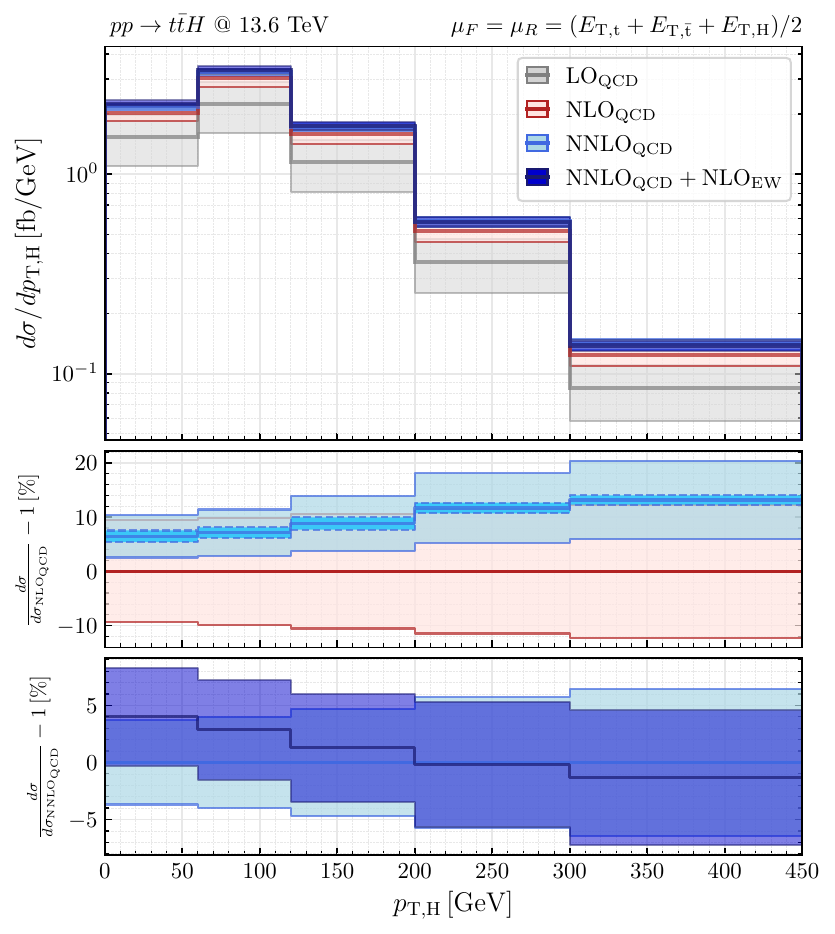}\hfill
     \includegraphics[height=0.38\textheight]{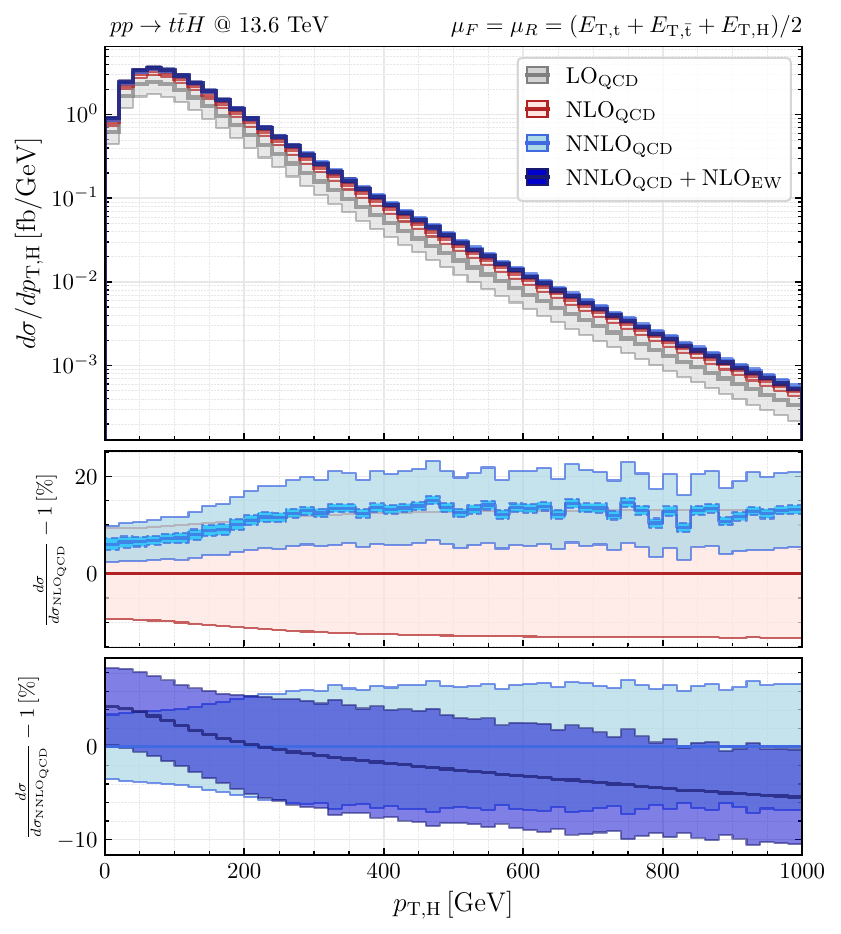}
  \caption{Differential cross section as a function of the Higgs boson transverse momentum, with the binning used by the LHC experimental collaborations~\cite{ATLAS:2021qou,CMS:2024fdo,ATLAS:2024gth} (on the left) and a finer binning (on the right). 
  The upper panel shows the $\mathrm{LO}_\mathrm{QCD}$ (grey), $\mathrm{NLO}_\mathrm{QCD}$ (red), $\mathrm{NNLO}_\mathrm{QCD}$ (light blue) and \mbox{$\mathrm{NNLO}_{\mathrm{QCD}}+\mathrm{NLO}_{\mathrm{EW}}$} (dark blue) curves, the central panel displays the $\mathrm{NLO}_\mathrm{QCD}$ and $\mathrm{NNLO}_\mathrm{QCD}$ predictions, relative to the former, and the lower panel shows the ratio to the $\mathrm{NNLO}_\mathrm{QCD}$ result, for the $\mathrm{NNLO}_\mathrm{QCD}$ and \mbox{$\mathrm{NNLO}_{\mathrm{QCD}}+\mathrm{NLO}_{\mathrm{EW}}$} curves. Bands indicate the (symmetrised) perturbative uncertainties, while the darker band in the central panel indicates the uncertainty in the $\mathrm{NNLO}_\mathrm{QCD}$ result stemming from the approximation of the two-loop virtual amplitudes.}
  \label{fig:pT_h}
\end{figure*}

We start by presenting the cross section as a function of the transverse momentum of the Higgs boson $p_\mathrm{T,H}$ in Fig.~\ref{fig:pT_h}.
In the left plot, we show results with the same binning used in the measurements performed by the LHC experimental collaborations and documented in Refs.~\cite{ATLAS:2021qou,CMS:2024fdo,ATLAS:2024gth}. To facilitate the use of our results, in Table~\ref{tab:pT_h} we provide the numerical values of the \mbox{$\mathrm{NNLO}_{\mathrm{QCD}}+\mathrm{NLO}_{\mathrm{EW}}$} cross section in the different $p_\mathrm{T,H}$ bins.

\begin{table}[b]
\centering
\renewcommand{\arraystretch}{1.5}
\setlength{\tabcolsep}{0.4em}
\begin{tabular}{lll}
$p_\mathrm{T,H}$ bin [GeV] & & $\sigma_{\mathrm{NNLO}_{\mathrm{QCD}}+\mathrm{NLO}_{\mathrm{EW}}}$ [fb] \\
 \toprule
$[0,60)$ & & 
$135.2(2)\phantom{0}\,^{+2.4\%}_{-4.1\%}\text{(scale)} \,{\scriptstyle\pm 1.0\%} \text{(approx)}$
\\
$[60,120)$ & & 
$200.2(2)\phantom{0}\,^{+2.3\%}_{-4.2\%}\text{(scale)} \,{\scriptstyle\pm 0.9\%} \text{(approx)}$
\\
$[120,200)$ & & 
$139.9(1)\phantom{0}\,^{+2.6\%}_{-4.7\%}\text{(scale)} \,{\scriptstyle\pm 1.0\%} \text{(approx)}$
\\
$[200,300)$ & & 
$\phantom{0}57.82(5)\,^{+3.4\%}_{-5.5\%}\text{(scale)} \,{\scriptstyle\pm 0.9\%} \text{(approx)}$
\\
$[300,450)$ & & 
$\phantom{0}20.76(3)\,^{+3.6\%}_{-6.0\%}\text{(scale)} \,{\scriptstyle\pm 0.8\%} \text{(approx)}$
\\
$[450,\infty)$ & & 
$\phantom{00}5.70(4)\,^{+3.4\%}_{-5.9\%}\text{(scale)} \,{\scriptstyle\pm 0.8\%} \text{(approx)}$
\\
\end{tabular}
\caption{\label{tab:pT_h} $t\bar{t}H$ production cross section in different bins of the Higgs boson transverse momentum. 
The first set of uncertainties (in percentage) indicates the customary 7-point scale variation, while the second set corresponds to the systematic uncertainty due to the approximation of the two-loop virtual corrections.
The combined uncertainties arising from the integration error and $q_T$-subtraction systematic uncertainty are indicated in parentheses.}
\end{table}

The $p_\mathrm{T,H}$ spectrum is shown with a finer binning in the right plot.
From the central panel, we observe a significant overlap between the $\mathrm{NNLO}_\mathrm{QCD}$ and $\mathrm{NLO}_\mathrm{QCD}$ bands, with the $\mathrm{NNLO}_\mathrm{QCD}$ uncertainties being roughly halved across the entire spectrum.
The NNLO QCD corrections also modify the shape of the distribution, being smaller at low $p_\mathrm{T,H}$, and gradually increasing until they reach a plateau at \mbox{$p_\mathrm{T,H}\sim 400\,\mathrm{GeV}$}.
In the lower panel, we can appreciate the effect of the EW corrections on top of the $\mathrm{NNLO}_\mathrm{QCD}$ result. The EW corrections are positive below \mbox{$p_\mathrm{T,H}\sim 200\,\mathrm{GeV}$} and contribute negatively in the high-$p_\mathrm{T,H}$ tail, thus inducing a quite significant shape distortion.
Their effects, however, turn out to be smaller than the residual perturbative uncertainties at $\mathrm{NNLO}_\mathrm{QCD}$ over the full range.
The uncertainties due to the approximate nature of the two-loop corrections, widely discussed in Sec.~\ref{sec:validation}, show that we achieve a very good control of our prediction over the whole spectrum, thanks to the complementarity between the SA and MA approaches in the different regions as well as for the relatively small overall impact of the approximated contribution.

\begin{figure*}[p]
  \includegraphics[height=0.38\textheight]{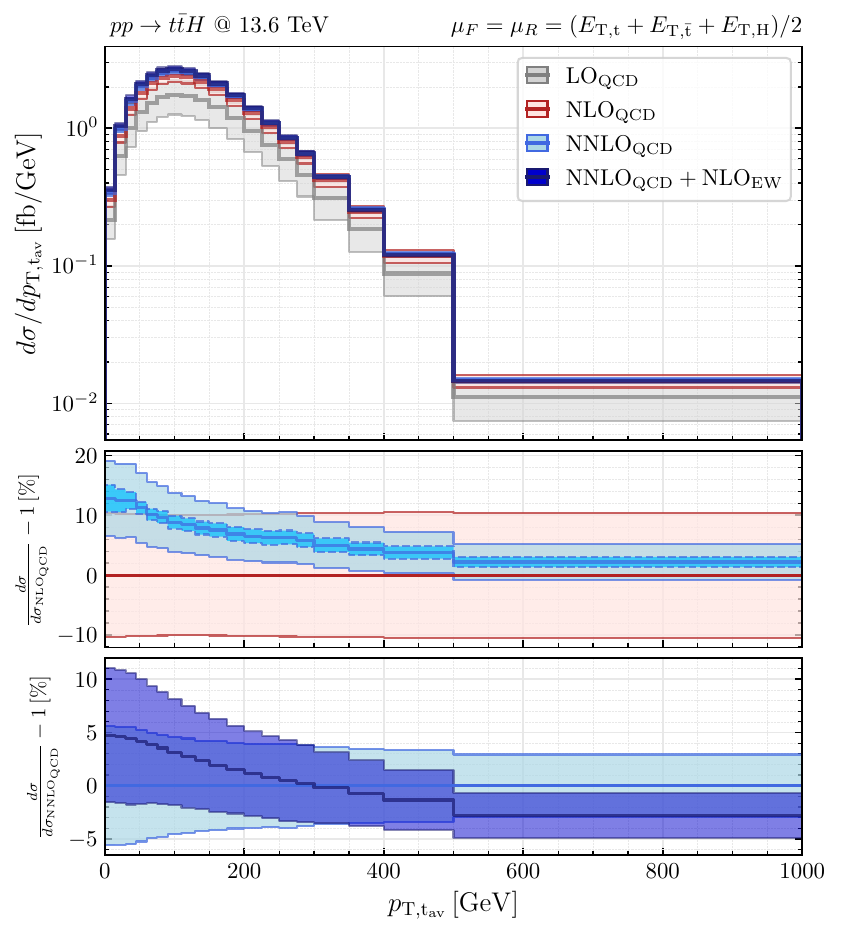}\hfill
  \includegraphics[height=0.38\textheight]{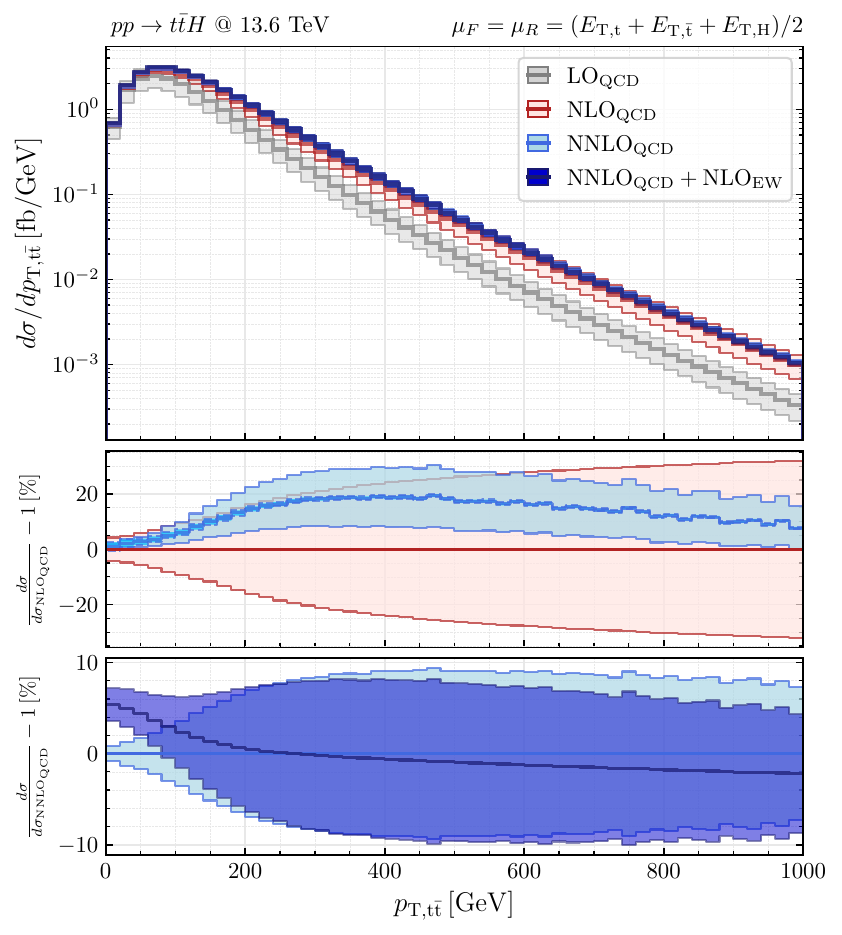}\\[4ex]
  \includegraphics[height=0.38\textheight]{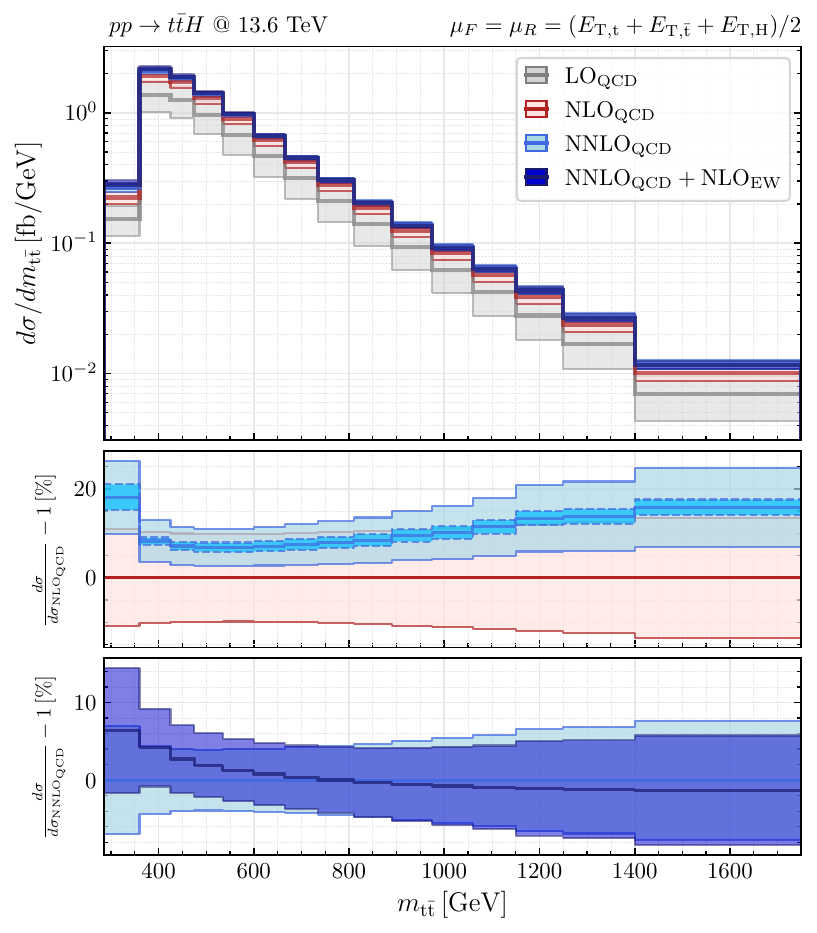}\hfill
  \includegraphics[height=0.38\textheight]{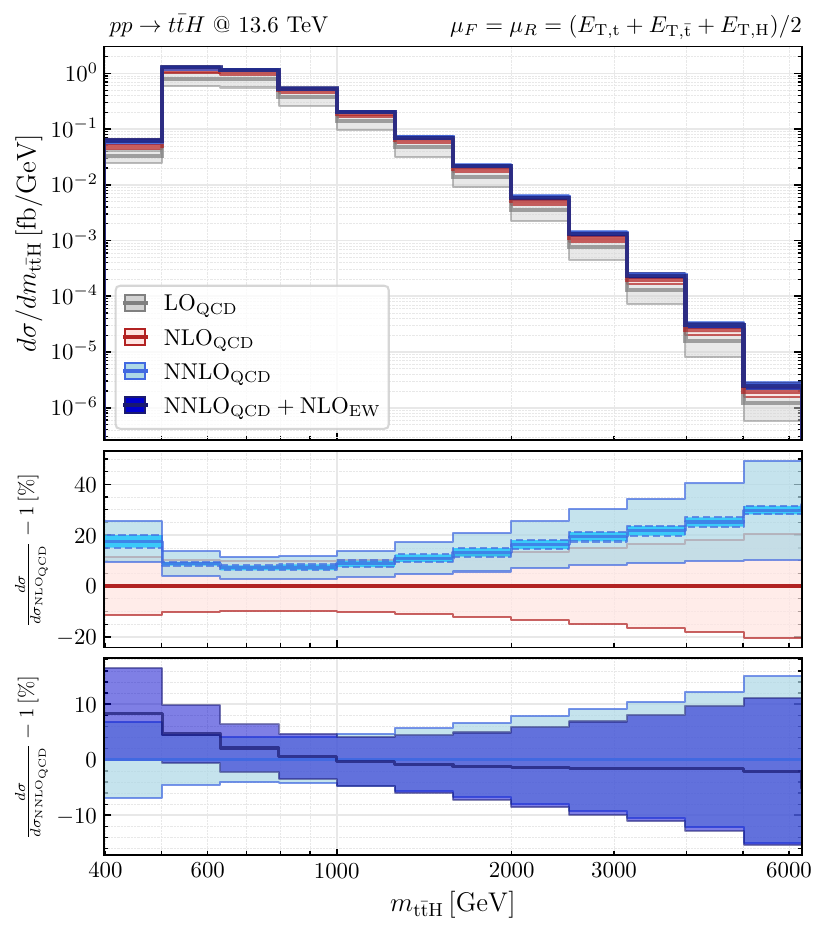}
  \caption{Differential cross section as a function of the average transverse momentum of the top quarks (top-left), the transverse momentum of the $t\bar{t}$ system (top-right) and its invariant mass (bottom-left), and the invariant mass of the $t\bar{t}H$ system (bottom-right). The layout is explained in the caption of Fig.~\ref{fig:pT_h}.}
    \label{fig:other_distributions}
\end{figure*}

In Fig.~\ref{fig:other_distributions} we show the transverse momentum spectrum of the average top quark, $p_{\mathrm{T,t}_\mathrm{av}}$, obtained by averaging the top quark and antiquark distributions.
Also in this case, the NNLO QCD corrections induce a significant change in the shape, with the corrections being more prominent at low transverse momentum. Their size considerably decreases with $p_{\mathrm{T,t}_\mathrm{av}}$, down to a few percent in the tail; this is also the region where scale uncertainties are reduced the most.
Similarly to the QCD pattern, the EW corrections are positive and larger for small transverse momentum, while they monotonically decrease with $p_{\mathrm{T,t}_\mathrm{av}}$ and eventually change sign. Once again, the prediction including EW corrections significantly overlaps with the $\mathrm{NNLO}_\mathrm{QCD}$ result.
The two-loop approximation uncertainties are also relatively uniform for this distribution, presenting only a slight increase at low $p_{\mathrm{T,t}_\mathrm{av}}$, while always being significantly smaller than the residual perturbative uncertainties.

Next, we discuss distributions for the top quark--antiquark system, namely its transverse momentum $p_\mathrm{T,t\bar{t}}$ and its invariant mass $m_\mathrm{t\bar{t}}$, also shown in Fig.~\ref{fig:other_distributions}.
Once again, the NNLO QCD corrections are not uniform across the spectrum. For the transverse momentum, the corrections are relatively small at low $p_\mathrm{T,t\bar{t}}$, increase to a maximum of about 20\% relative to the NLO at \mbox{$p_\mathrm{T,t\bar{t}}\sim 350\,\mathrm{GeV}$}, and then decrease again to $5-10\%$ at \mbox{$p_\mathrm{T,t\bar{t}}\sim 1000\,\mathrm{GeV}$}. In contrast, for the invariant mass, the trend is reversed: the corrections are significant at the threshold, decrease to a minimum around \mbox{$m_\mathrm{t\bar{t}} \sim 500\,\mathrm{GeV}$}, and then increase again in the high-mass tail. 
For both distributions, the scale uncertainties are significantly reduced at NNLO, and there is always a sizeable overlap with the NLO prediction, the only exception being the $m_\mathrm{t\bar{t}}$ threshold where the overlap is marginal.
The EW corrections are moderate in the tails, while they become negative and larger than 5\% for both low $p_\mathrm{T,t\bar{t}}$ and low $m_\mathrm{t\bar{t}}$. 
For \mbox{$p_\mathrm{T,t\bar{t}} < 50\,\mathrm{GeV}$} we observe no overlap between the $\mathrm{NNLO}_\mathrm{QCD}$ and \mbox{$\mathrm{NNLO}_{\mathrm{QCD}}+\mathrm{NLO}_{\mathrm{EW}}$} bands, while the overlap is very sizeable anywhere else.
The QCD perturbative uncertainties are relatively small in this region both at NLO and NNLO. Still, since the NNLO band is entirely contained in the NLO one, we do not expect these uncertainties to be substantially underestimated. 
The two-loop approximation uncertainties are negligible in the tail of the $p_\mathrm{T,t\bar{t}}$ distribution, while they become similar in size to the scale uncertainties at low transverse momentum. 
In the case of $m_\mathrm{t\bar{t}}$, these uncertainties are more significant both at low and high invariant masses, always subdominant though when compared to scale variations.
In Fig.~\ref{fig:other_distributions}, we also present the invariant mass of the entire $t\bar{t}H$ system. The pattern of QCD and EW corrections is quite similar to that observed for $m_\mathrm{t\bar{t}}$.

\begin{figure*}[t]
  \includegraphics[height=0.38\textheight]{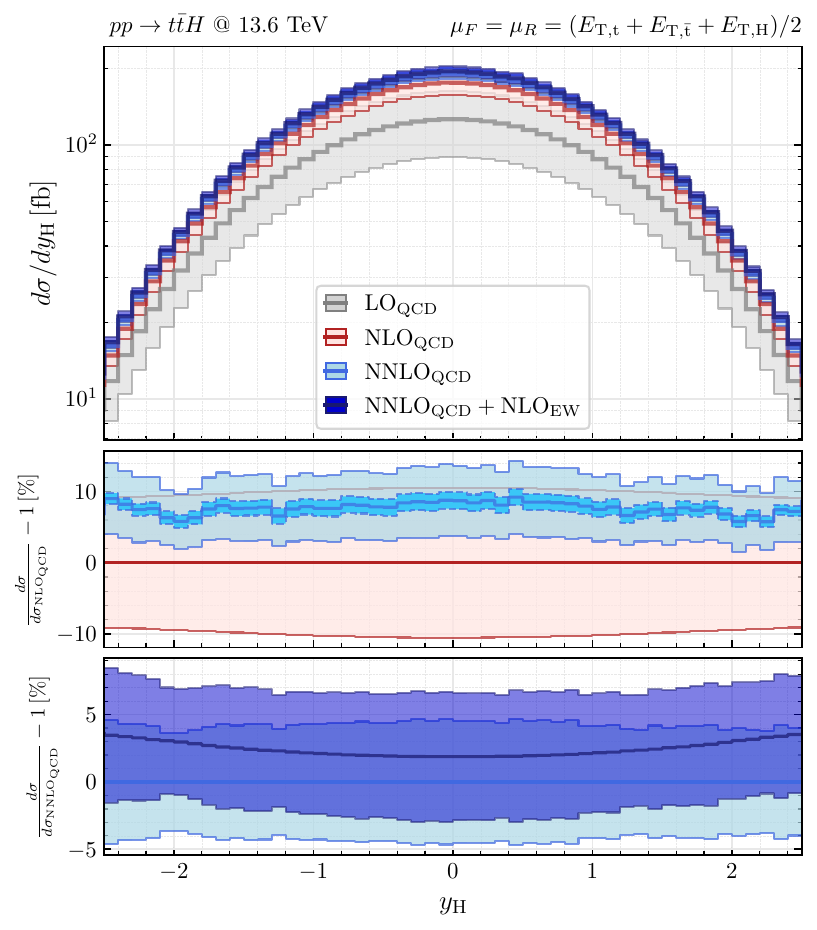}\hfill
  \includegraphics[height=0.38\textheight]{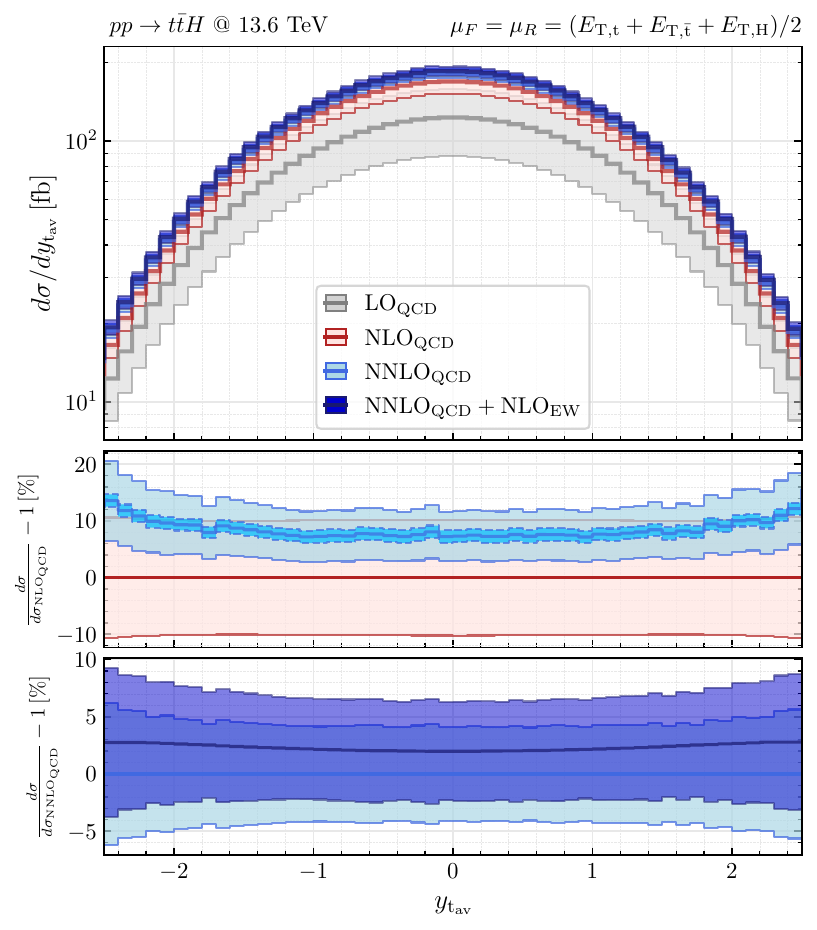}
  \caption{Differential cross section as a function of the Higgs boson (left) and the average top-quark (right) rapidities. The layout is explained in the caption of Fig.~\ref{fig:pT_h}.}
    \label{fig:rapidities}
\end{figure*}

Finally, we discuss the rapidity distributions of the Higgs boson and the average top quark, shown in Fig.~\ref{fig:rapidities}.
The QCD corrections turn out to be rather flat in this case, the only exception being the large-$|y_{\mathrm{t}_\mathrm{av}}|$ regions where the NNLO corrections become larger.
Including the whole tower of EW corrections leads to a few-percent increase with respect to the NNLO cross section, which is quite constant and tends to be smaller in the central-rapidity region. This pattern is more evident for $y_\mathrm{H}$.
The uncertainties arising from the approximation of the NNLO virtual corrections are rather flat over the rapidity spectra and significantly smaller than perturbative uncertainties.

\section{Summary}
\label{sec:summa}

In this work, we have presented the first fully differential NNLO QCD calculation for \ttH production at hadron colliders.
Our calculation is exact, except for the finite remainder of the two-loop virtual corrections, whose exact evaluation still appears beyond current possibilities.
To estimate their size, we relied on two distinct approaches: a soft Higgs boson approximation and a high-energy expansion in the small top-mass limit. 
The former relies on a factorisation formula we developed and proved up to the three-loop order. The latter is based on a well-established technique known as massification.
The two approximations are formally valid in distinct dynamical regions and, moreover, are constructed using completely different perturbative ingredients: in one case, the two-loop amplitudes for $t\bar{t}$ production \cite{Barnreuther:2013qvf}, and in the other, those for a Higgs boson plus four massless partons \cite{Badger:2021ega}.
Despite this, both approaches yield consistent results within their respective uncertainties, providing strong evidence of the robustness of our method. 

We have combined these two approaches to derive our final results and developed a method to conservatively estimate the associated systematic uncertainties for both the total cross section and differential distributions.
Thanks to the relatively small impact of the pure two-loop contribution, we obtain NNLO QCD predictions with a systematic uncertainty of $\mathcal{O}(1\%)$ at the inclusive level and, therefore, much smaller than the residual perturbative uncertainties.
This statement holds true also at the differential level, except for a few specific regions in phase space. 
In particular, our NNLO predictions for the transverse momentum of the Higgs boson, currently measured by the ATLAS and CMS collaborations, are affected by a systematic uncertainty due to the approximation of the two-loop virtual corrections, which remains $\mathcal{O}(1\%)$ in every bin of the distribution.

Finally, we have equipped our NNLO QCD results with the complete tower of NLO EW corrections, thus obtaining the most accurate predictions available to date for this process. Perturbative uncertainties at \mbox{$\mathrm{NNLO}_{\mathrm{QCD}}+\mathrm{NLO}_{\mathrm{EW}}$} are at the level of 3\% for the total cross section, and about $4-6\%$ for the $p_\mathrm{T,H}$ spectrum.
We are confident that our predictions will contribute to fully exploit the potential of the upcoming $\ttH$ measurements performed at the LHC and HL-LHC, particularly if a good control of the backgrounds will be achieved.

\vspace*{0.7cm}
\noindent {\bf Acknowledgements}    

\noindent We would like to thank Kay Sch\"onwald for his help in extracting the soft limit of the three-loop scalar form factor from the expressions provided in Ref.~\cite{Fael:2022miw}, and Christian Biello for extensive cross-checks on the implementation of the massless $b \bar bH$ amplitudes.
We are grateful to Pia Bredt for the validation of our EW predictions against her new implementation in \textsc{Whizard}, and to Jonas Lindert for his ongoing support on {\sc OpenLoops}, in particular for providing several dedicated amplitudes.
We are indebted to Christopher Schwan for his immediate support on {\sc PineAPPL}, including explicit extensions of its functionality.
We also thank Vasily Sotnikov and Simone Zoia for useful discussions.
This work is supported in part by the Swiss National Science Foundation (SNSF) under contract 200020$\_$219367. 
The work of S.D. has been funded by the European Union (ERC, MultiScaleAmp, Grant Agreement No. 101078449). Views and opinions expressed are however those of the author(s) only and do not necessarily reflect those of the European Union or the European Research Council Executive Agency. 
Neither the European Union nor the granting authority can be held responsible for them.

\paragraph{Note added.}
After the submission of this paper, Ref.~\cite{Badger:2024awe} appeared where the two-loop $\mathcal{Q}\overline{\mathcal{Q}} H$ massless amplitudes, an essential ingredient to construct our high-energy approximation, became available in full colour. Incorporating the massless subleading-colour contributions into our framework leads to a reduction of the hard-virtual contribution by $\mathcal{O}(30\%)$ at high $p_{\mathrm{T,H}}$, along with a significant decrease in the $\muIR$-variation error. The overall impact on our NNLO results at high $p_{\mathrm{T,H}}$ appears to be within our estimated systematic uncertainties.

\appendix

\section{Soft-Higgs factorisation at \texorpdfstring{N$^3$LO}{N3LO}}
\label{app:soft}
The relation in Eq.~\eqref{eq:deltaHQQ-F} is expected to be valid at all perturbative orders.
This appendix explicitly shows its validity up to the three-loop order (N$^3$LO) in QCD.
This is possible since the exact computation of the three-loop massive scalar form factor, including both singlet and non-singlet contributions, has recently been made available in Ref.~\cite{Fael:2022miw}.

By applying the LET, the first step consists of computing the right-hand side of Eq.~\eqref{eq:LET_selfenergy} up to three-loop order, namely
\begin{align}
	m_0 \frac{\partial }{\partial m_0}\M_{\mathcal{Q} \to \mathcal{Q}}^{\mathrm{bare}}(p) \biggl |_{p^2 = m^2} = -m_0 \overline{\mathcal{Q}}_{0} \mathcal{Q}_{0} \biggl \{ 1 &+  
	\sum_{n=1}^{3}\,\biggl[ \frac{g_0^2}{(4\pi)^{d/2}m^{2\epsilon}} \biggr]^n \,\biggl[ A_n(Z_m^2) - B_n(Z_m^2) \biggr]     \notag \\
	 &+ 2 Z_m^2 \sum_{n=1}^{3} \,\biggl[ \frac{g_0^2}{(4\pi)^{d/2}m^{2\epsilon}} \biggr]^n \,\frac{\partial}{\partial t} \biggl[ A_n(t) - B_n(t) \biggr] \biggl |_{t= Z_m^2}     \notag \\
	 &+ 2 Z_m \sum_{n=1}^{3} \,\biggl[ \frac{g_0^2}{(4\pi)^{d/2}m^{2\epsilon}} \biggr]^n \,\frac{\partial}{\partial t} B_n(t) \biggl |_{t= Z_m^2} + \mathcal{O}(g_0^8)
	 \biggr\} \,,
\end{align}
where \mbox{$t \equiv m_0^2/p^2$}. After expanding the previous equation around \mbox{$Z_m^2 = 1$} (where $Z_m$ is given in Eq.~\eqref{eq:Z_m}) and neglecting contributions of order $\mathcal{O}(g_0^8)$, we find that the $\mathcal{O}(g_0^6)$ correction to Eq.~\eqref{eq:bare_amplitude} is 
\begingroup
\allowdisplaybreaks
\begin{align}
	\hspace{-0.9cm} 
	\left(\! m_0 \frac{\partial }{\partial m_0}\M_{\mathcal{Q} \to \mathcal{Q}}^{\mathrm{bare}}(p) \biggl |_{p^2 = m^2} \!\right)_{\mathcal{O}(g_0^6)} \!\!\!\!\!=  
	-m_0 \overline{\mathcal{Q}}_{0} \mathcal{Q}_{0} \left( \frac{g_0^2}{(4\pi)^{d/2}m^{2\epsilon}} \right)^3
	&\! \biggl \{ 
		A_3 - B_3 + 2A^{'}_3 + 2 M_1 (2 A^{''}_2 + 3 A^{'}_2 -2 B^{'}_2 )  \notag \\
		& + M_1^2 (4 A^{'''}_1 + 12 A^{''}_1 + 3 A^{'}_1 -6 B^{''}_1 - 3 B^{'}_1)  \notag \\
		& + 2 M_2 (2 A^{''}_1 + 3 A^{'}_1 -2 B^{'}_1) 
	\biggr\} \,,
	 \label{eq:three-loop_bare_amplitude}
\end{align}
\endgroup
where all coefficients $A_n, B_n$ and their derivatives are evaluated at \mbox{$p^2=m_0^2$}. 

The previous result cannot be directly used since the three-loop analytic expression of the quark self-energy is not available in the literature to the best of our knowledge. 
This implies that we have to find a relation between the coefficients $A_3, B_3$, the higher derivatives of the coefficients $A_2,A_1,B_2,B_1$ and the three-loop expression of the on-shell wave-function and mass renormalisation constants. 
In the on-shell scheme, the three-loop contribution to $Z_2$ and $Z_m$ has been computed and can be extracted, for example, from Ref.~\cite{Melnikov:2000zc}.

We start considering the renormalised quark propagator
\begin{equation}
	\frac{Z_2}{\slashed{p} - m} = \frac{1}{\slashed{p} - m_0 + \Sigma(\slashed{p},m_0)}
\end{equation} 
and expanding the denominator in the r.h.s.\ around the on-shell mass condition \mbox{$p^2=m^2$}. 
By identifying the pole with the pole mass and the residue with the wave-function renormalisation constant, we obtain the following crucial (all-order) relations
\begingroup
\allowdisplaybreaks
\begin{align}
	Z_m -1 &= Z_m \Sigma_S(p^2=m^2, Z_m^2) + \Sigma_V(p^2=m^2, Z_m^2) \,,  \\
	1/Z_2 &= 1 + \Sigma_V(p^2=m^2, Z_m^2)  \notag \\
	&+ 2\epsilon Z_m \sum_{n=1}^{+\infty}\,\biggl[ \frac{g_0^2}{(4\pi)^{d/2}m^{2\epsilon}} \biggr]^n \, n A_n(Z_m^2) 
	+ 2\epsilon (1-Z_m) \sum_{n=1}^{+\infty}\,\biggl[ \frac{g_0^2}{(4\pi)^{d/2}m^{2\epsilon}} \biggr]^n \, n B_n(Z_m^2)  \notag \\
	&+ 2Z_m^3 \sum_{n=1}^{+\infty}\,\biggl[ \frac{g_0^2}{(4\pi)^{d/2}m^{2\epsilon}} \biggr]^n A^{'}_n(Z_m^2)
	+ 2Z_m^2(1-Z_m) \sum_{n=1}^{+\infty}\,\biggl[ \frac{g_0^2}{(4\pi)^{d/2}m^{2\epsilon}} \biggr]^n B^{'}_n(Z_m^2)  \,.
\end{align}
\endgroup
The expansion of both sides of the first equation in the strong coupling leads to
\begin{align}
	A_1 &= -M_1 \\
	A_2 &= -M_2 - M_1(A_1-B_1 +2A^{'}_1) \\
	A_3 &= -M_3 - M_2(A_1-B_1 +2A^{'}_1) -M_1(A_2 -B_2 + 2A^{'}_2) -M_1^2(2A^{''}_1+3A^{'}_1-2B^{'}_1) \,,
\end{align}
while, from the second equation, we can extract a relation for the first derivatives
\begin{align}
	A^{'}_1 &= \frac{1}{2} ( B_1 - F_1) - \epsilon A_1 \\
	A^{'}_2 &= \frac{1}{2} ( B_2 - F_2) - 2 \epsilon A_2 + \frac{(d^2 - 7d + 8 + a_0)}{2 (d - 5)} A_1^2  \\
	A^{'}_3 &= \frac{1}{2} \biggl[ B_3 -  F_3  - F_1^3 + 2 F_1 F_2 
	- 2 M_2(2 A^{''}_1 + 3 A^{'}_1 - 2 B^{'}_1) \notag \\ 
	& - 2 M_1(2 A^{''}_2  + 3 A^{'}_2 - 2 B^{'}_2)
	- M_1^2(4 A^{'''}_1 + 14 A^{''}_1 + 6 A^{'}_1 - 6 B^{''}_1 - 5 B^{'}_1)   \notag  \\
	&- 2 \epsilon \biggl( 3 A_3  + 2 M_1(A_2 - B_2 + 2 A^{'}_2)
	+ M_1^2(2 A^{''}_1 + 3 A^{'}_1 - 2 B^{'}_1)  
	+ M_2( A_1 - B_1 +2 A^{'}_1) \biggr) \biggr]  \,,
\end{align}
where $a_0$ denotes the gauge parameter.

After inserting the previous relations in Eq.~\eqref{eq:three-loop_bare_amplitude} and renormalising the heavy-quark mass and wave function, we find that the coefficient of the $\mathcal{O}(g_0^6)$ contribution reads
\begin{equation}
	a_0 A_1^3 + A_1^2 (4 A^{''}_1 + 3B_1 -4 B^{'}_1)(1+2 \epsilon) + 6 M_3 \epsilon + 2 A_1 A_2 \epsilon(-3 + 8\epsilon) + 2A_1^3\epsilon(7 -20\epsilon +8\epsilon^2) \,.
\end{equation}
We note that all $A_n, B_n$ terms present in the previous formula can be taken from Refs.~\cite{Gray:1990yh,Broadhurst:1991fy}, except for the three-loop correction to $Z_m$ (i.e.\ $M_3$). The analytic expression of this coefficient can be extracted from Ref.~\cite{Melnikov:2000zc} up to a change in the overall normalisation.

Finally, we renormalise the strong coupling in the $\MSbar$ scheme (with \mbox{$n_f = n_l+n_h$} active flavours) via the two-loop relation
\begingroup
\allowdisplaybreaks
\begin{equation}
	\frac{g_0^2}{4\pi} = \biggl(\frac{e^{\gamma_E} \mu^2}{4\pi} \biggr)^{\!\epsilon} \alphasNf(\mu^2) \biggl\{ 1- \frac{\alphasNf(\mu^2) }{2 \pi} \frac{\beta_0^{(n_f)}}{\epsilon} 
	+ \biggl( \frac{\alphasNf(\mu^2) }{2 \pi} \biggr)^2 \biggl[ \biggl(\frac{\beta_0^{(n_f)}}{\epsilon}\biggr)^2  - \frac{\beta_1^{(n_f)}}{2\epsilon}\biggr] + \mathcal{O}(\as^3) \biggr\} \,,
\end{equation} 
\endgroup
where $\beta_0^{(n_f)}$ is given in Eq.~\eqref{eq:first-coeff_betafunction} (by replacing \mbox{$n_l \to n_f$}) and 
\begin{equation}
	\beta_1^{(n_f)} = \frac{17 C_A^2 -(5 C_A + 3C_F) n_f}{6} \,.
	\label{eq:second-coeff_betafunction}
\end{equation}
We find that the three-loop contribution \mbox{$\left(\alphasNf(\mu^2)/ \pi\right)^{\!3} \delta_{H \mathcal{Q}\bar{\mathcal{Q}}}^{(3)}$} to $\delta_{H\mathcal{Q}\bar{\mathcal{Q}}}$, before applying the decoupling of $n_h$ heavy quarks of mass $m$, is
\begingroup
\allowdisplaybreaks
\begin{align}
	\delta_{H\mathcal{Q}\bar{\mathcal{Q}}}^{(3)} &= 
	C_A^2 C_F \biggl[ -\frac{121}{96} \ln ^2\left(\frac{\mu ^2}{m^2}\right)-\frac{2341}{288} \ln
   	\left(\frac{\mu ^2}{m^2}\right) -\frac{13243}{864} +\frac{11}{6} \zeta_2 +\frac{11 \zeta_3}{8} -\frac{11}{2} \zeta_2 \ln (2) \biggr]  \notag \\
	&+ C_A C_F^2 \biggl[ \frac{121}{32} \ln \left(\frac{\mu ^2}{m^2}\right) +\frac{175}{16} -\frac{55}{8} \zeta_2 -\frac{11 \zeta_3}{4} + 11\zeta_2 \ln (2) \biggr]  \notag \\
	& - C_F^3  \frac{309}{64}  \notag \\
	& + C_F C_A n_h \biggl[ \frac{11}{24} \ln ^2\left(\frac{\mu ^2}{m^2}\right)+\frac{373}{144} \ln
   \left(\frac{\mu ^2}{m^2}\right) +\frac{583}{108} -\frac{13}{6} \zeta_2 +\frac{\zeta_3}{2} + \zeta_2 \ln (2) \biggr]  \notag \\
   	& + C_F^2 n_h \biggl[ -\frac{1}{2} \ln \left(\frac{\mu ^2}{m^2}\right)-\frac{23}{64} +\frac{5}{4}\zeta_2 -\frac{\zeta_3}{4} -2 \zeta_2\ln (2) \biggr]  \notag \\
   	& + C_F n_h^2 \biggl[ -\frac{1}{24} \ln ^2\left(\frac{\mu ^2}{m^2}\right)-\frac{13}{72} \ln
   \left(\frac{\mu ^2}{m^2}\right) -\frac{197}{432} +\frac{\zeta_2}{3}  \biggr]  \notag \\
   	& + C_F C_A n_l \biggl[ \frac{11}{24} \ln ^2\left(\frac{\mu ^2}{m^2}\right)+\frac{373}{144} \ln
   \left(\frac{\mu ^2}{m^2}\right)+\frac{869}{216} +\frac{7}{12} \zeta_2  +\frac{\zeta_3}{2}+ \zeta_2\ln (2) \biggr]  \notag \\
   	& + C_F^2 n_l \biggl[ -\frac{1}{2} \ln \left(\frac{\mu ^2}{m^2}\right) -\frac{23}{64} +\frac{5}{4}\zeta_2 -\frac{\zeta_3}{4} - 2\zeta_2 \ln (2)  \biggr]  \notag \\
   	& + C_F n_l^2 \bigl[ -\frac{1}{24} \ln ^2\left(\frac{\mu ^2}{m^2}\right)-\frac{13}{72} \ln
   \left(\frac{\mu ^2}{m^2}\right) -\frac{89}{432} -\frac{\zeta_2}{6}  \biggr] \notag \\
   	& + C_F n_l n_h \biggl[ -\frac{1}{12} \ln ^2\left(\frac{\mu ^2}{m^2}\right)-\frac{13}{36} \ln
   \left(\frac{\mu ^2}{m^2}\right) -\frac{143}{216}  +\frac{\zeta_2}{6}\biggr] \,.
   \label{eq:deltaHQQ_three-loop}
\end{align}
\endgroup
As expected, the three-loop coefficient $\delta_{H\mathcal{Q}\bar{\mathcal{Q}}}^{(3)}$ is IR finite (all $\epsilon$ poles cancel out) and gauge-independent (no residual dependence on the gauge parameter $a_0$). This is a nontrivial check of the correctness of our result. 
Even though the complete expression of the soft limit of the three-loop massive scalar form factor in the on-shell scheme is not provided in Ref.~\cite{Fael:2022miw}, we obtained it from one of the authors. We find perfect agreement with our result in Eq.~\eqref{eq:deltaHQQ_three-loop}, derived using the Higgs LET, thereby confirming the validity of Eqs.~\eqref{eq:fact} and \eqref{eq:deltaHQQ-F} at three-loop order.

The final step implies the decoupling of $n_h$ heavy quarks from the running of $\as$. 
After applying the two-loop decoupling relation, in the on-shell mass scheme (see e.g.\ Ref.~\cite{Chetyrkin:1997un}),
\begingroup
\allowdisplaybreaks
\begin{align}
	\alphasNf(\mu^2) &= \alphasNl(\mu^2) \biggl\{ 1 + \frac{\alphasNl(\mu^2) }{2\pi}\frac{n_h }{3} \ln \left(\frac{\mu ^2}{m^2}\right) \notag \\
	&+ \biggl( \frac{\alphasNl(\mu^2) }{2 \pi} \biggr)^{\!2} n_h \left[ \frac{15}{8}C_F -\frac{4}{9}C_A +\biggl( \frac{C_F}{2} + \frac{5}{6} C_A  \biggr)\ln \left(\frac{\mu ^2}{m^2}\right)  
	+ \frac{n_h}{9} \ln^2 \left(\frac{\mu ^2}{m^2}\right)   \right]
	+ \mathcal{O}(\as^3) \biggr\} \,,
	\label{eq:twoloop_decoupling}
\end{align}
\endgroup
we extract the correction to the Higgs--quark coupling in the soft-Higgs limit up to $\mathcal{O}(\as^3)$,
\begingroup
\allowdisplaybreaks
\begin{align}
\hspace{-0.9cm}
	\delta_{H \mathcal{Q}\bar{\mathcal{Q}} } &= \frac{\alphasNl(\mu^2)}{2 \pi} \left(-3 C_F\right) \notag \\
  &+\biggl(\! \frac{\alphasNl(\mu^2)}{2 \pi} \!\biggr)^2 \biggl\{ \frac{33}{4}C_F^2-\frac{185}{12} C_F C_A+\frac{13}{6} C_F (n_l+n_h)- 3C_F\beta^{(n_l)}_0\ln\left(\! \frac{\mu^2}{m^2} \!\right) \biggr\} \notag \\
  &+ \biggl(\! \frac{\alphasNl(\mu^2) }{2 \pi} \! \biggr)^3 \! \biggl\{
	C_A^2 C_F \biggl[ -\frac{121}{12} \ln^2\left(\! \frac{\mu ^2}{m^2} \!\right)-\frac{2341}{36} \ln
   	\left(\! \frac{\mu ^2}{m^2} \!\right) -\frac{13243}{108} +\frac{44}{3}\zeta_2 + 11 \zeta_3  -44 \zeta_2 \ln (2) \biggr]  \notag \\
	&\hspace{2.3cm} + C_A C_F^2 \biggl[ \frac{121}{4} \ln \left(\! \frac{\mu ^2}{m^2} \! \right) +\frac{175}{2} - 55\zeta_2 -22\zeta_3 + 88 \zeta_2 \ln (2) \biggr]  \notag \\
	&\hspace{2.3cm} - C_F^3  \frac{309}{8}  \notag \\
	&\hspace{2.3cm} + C_F C_A n_h \biggl[ 
	\frac{143}{18} \ln \left(\! \frac{\mu ^2}{m^2} \!\right) +\frac{1202}{27} -\frac{52}{3} \zeta_2 +4\zeta_3 + 8 \zeta_2 \ln (2)
	\biggr]  \notag \\
	&\hspace{2.3cm} + C_F^2 n_h \biggl[ -\frac{17}{2} + 10\zeta_2 -2\zeta_3 - 16 \zeta_2 \ln (2) \biggr]  \notag \\
   	&\hspace{2.3cm} + C_F n_h^2 \biggl[ -\frac{197}{54} + \frac{8}{3}\zeta_2  \biggr]  \notag \\
   	&\hspace{2.3cm} + C_F C_A n_l \biggl[ \frac{11}{3} \ln ^2\left(\! \frac{\mu ^2}{m^2} \!\right)+\frac{373}{18} \ln
   \left(\! \frac{\mu ^2}{m^2} \!\right) + \frac{869}{27} +\frac{14}{3}\zeta_2 +4\zeta_3 + 8 \zeta_2 \ln (2) \biggr]  \notag \\
   	&\hspace{2.3cm} + C_F^2 n_l \biggl[ -4 \ln \left(\! \frac{\mu ^2}{m^2} \!\right) -\frac{23}{8} + 10\zeta_2  -2\zeta_3 - 16 \zeta_2 \ln (2)  \biggr]  \notag \\
   	&\hspace{2.3cm} + C_F n_l^2 \biggl[ -\frac{1}{3} \ln ^2\left(\! \frac{\mu ^2}{m^2} \!\right)-\frac{13}{9} \ln
   \left(\! \frac{\mu ^2}{m^2} \!\right) -\frac{89}{54} -\frac{4}{3} \zeta_2  \biggr] \notag \\
   	&\hspace{2.3cm} + C_F n_l n_h \biggl[ -\frac{13}{9} \ln \left(\! \frac{\mu ^2}{m^2} \!\right) -\frac{143}{27} +\frac{4}{3}\zeta_2 \biggr]
	\biggr\} +{\cal O}(\as^4) \,.
\end{align}
\endgroup
Unlike the two-loop result, logarithmic contributions in $m$ dependent on the number of heavy-quark flavours $n_h$ persist at three-loop order.
However, a pattern emerges: the coefficient of \mbox{$\ln \left(\! \frac{\mu ^2}{m^2} \!\right)$} at three-loop order is \mbox{$\beta_0^{(n_l)} \frac{13}{24} C_F n_h$}, which corresponds to \mbox{$\beta_0^{(n_l)}$} times the two-loop $n_h$-dependent coefficient \mbox{$\frac{13}{24} C_F n_h$}, in $\frac{\as}{\pi}$ expansion.

\section{Momentum mappings}
\label{app:mapping}

In this appendix, we describe the mappings used to implement the high-energy approximation of Sect.~\ref{subsubsec:massification_HQQ}.

We start from a set of momenta \mbox{$\{p_1,p_2,p_3,p_4,q\}$}, according to the ordering in Eq.~\eqref{HQQ_process}, with \mbox{$p_3^2=p_4^2=m^2$} and \mbox{$q^2=m_H^2$}.
In order to evaluate the massive finite remainder in Eq.~\eqref{master_equation5} we need to define a corresponding set of momenta \mbox{$\{\tilde{p}_1,\tilde{p}_2,\tilde{p}_3,\tilde{p}_4,q\}$} with \mbox{$\tilde{p}^2_3=\tilde{p}^2_4=0$}. 
The two sets of momenta should only differ by power corrections in the heavy-quark mass $m$. 
We explored the following three projections:
\begin{itemize}
 	\item \underline{mapping 0}:
	As a first attempt, we preserve the four-momenta of the incoming partons and the Higgs boson. As a consequence, also the four-momentum of the heavy-quark pair is unchanged, i.e.\ in particular 
	\begin{equation}
		s_{\mathcal{Q}\mathcal{Q}} = (p_3 + p_4)^2 = (\tilde{p}_3 + \tilde{p}_4)^2 \,.
	\end{equation}
	The new massless momenta $\tilde{p}_3$ and $\tilde{p}_4$ are defined as linear combinations of the original $p_3$ and $p_4$,
	\begingroup
	\allowdisplaybreaks
	\begin{align}
		\tilde{p}_3^{\mu} &= \beta_{+}p_3^{\mu} - \beta_{-}p_4^{\mu} \,, \notag \\
		\tilde{p}_4^{\mu} &= \beta_{+}p_4^{\mu} - \beta_{-}p_3^{\mu} \,,
	\end{align}
	\endgroup
	with \mbox{$\beta_{\pm} = \frac{1\pm \beta}{2 \beta}$}, \mbox{$\beta = \sqrt{1-\frac{4m^2}{s_{\mathcal{Q}\mathcal{Q}}}}$}.
 	Potential divergences related to a collinear splitting \mbox{${g\to\mathcal{Q}\mathcal{Q}}$} in the massless matrix elements are avoided by construction.
        This mapping was successfully used in the NNLO computations of $b\bar b W$ \cite{Buonocore:2022pqq} and $t\bar t W$ \cite{Buonocore:2023ljm} and
        turns out to work nicely also for \mbox{${q\bar{q} \to b \bar b H}$},
        but is not ideal for \ttH production or, in other words, when we increase the mass of the heavy quark. 
	In the latter case, the approximated one-loop hard-virtual contribution based on this mapping differs by more than a factor of two from the exact one-loop finite remainder.
	\item \underline{mapping 1}:
	The second possibility we consider is a mapping that
	\begin{itemize}
		\item preserves the four-momentum $q$ of the Higgs boson; 
		\item keeps  the tri-momenta of the heavy quarks unchanged while rescaling their energy to ensure the massless on-shell condition;
		\item accordingly modifies the momenta of the initial-state partons to guarantee momentum conservation.
	\end{itemize}
	To be precise, the projected momenta are defined as
	\begin{align}
		\tilde{p}_1^{\mu} &= \tilde{E}_1(1,0,0,1), ~~~ \tilde{p}_2^{\mu} = \tilde{E}_2(1,0,0,-1) \,,\notag \\
		\tilde{p}_3^{\mu} &= (\sqrt{p_{3,T}^2+p_{3,z}^2}, \vec{p}_{3,T}, p_{3,z})  \,,\notag \\
		\tilde{p}_4^{\mu} &= (\sqrt{p_{4,T}^2+p_{4,z}^2}, \vec{p}_{4,T}, p_{4,z})  \,,
	\end{align}
	with \mbox{$\tilde{E}_1 = \frac{1}{2}\biggl(q^0+q_z+\displaystyle\sum_{i=3,4}\left(\tilde{E}_i+ p_{i,z}\right)\biggr)$} and 
	\mbox{$\tilde{E}_2 = \frac{1}{2}\biggl(q^0-q_z+\displaystyle\sum_{i=3,4}\left(\tilde{E}_i- p_{i,z}\right)\biggr)$}.
	\item \underline{mapping 2}:
	An undesirable side effect of the previous two mappings is that they can lead to projected events where one of the heavy quarks is collinear to one of the initial-state momenta. 
	This represents an issue for the $gg$ channel since, in this limit, the massless amplitudes are divergent, and the heavy-quark mass no longer screens the singularity.
	Therefore, we introduce a third mapping that avoids these problematic configurations. 
	The idea is to prevent the transverse momentum of the projected heavy quarks from being smaller than a technical cutoff of the order of the heavy-quark mass $m$.
More precisely:
	\begin{itemize}
		\item we preserve the energy and longitudinal component of the four-momenta of the heavy quarks and rescale their transverse momentum to ensure the on-shell massless condition,
		\begin{align}
			\tilde{p}_3^{\mu} &= (E_3, \hat{p}_{3,T} \sqrt{p_{3,T}^2 + m^2}, p_{3,z}) \,, \notag \\
			\tilde{p}_4^{\mu} &= (E_4, \hat{p}_{4,T} \sqrt{p_{4,T}^2 + m^2}, p_{4,z}) \,,
		\end{align}
		where \mbox{$\hat{p}_{i,T} = {\vec{p}_{i,T}}/{p_{i,T}}$} for \mbox{$i\in\{3,4\}$};
		\item we restore momentum conservation by redistributing \mbox{$\Delta p_{34} = p_3 + p_4 - \tilde{p}_3 -\tilde{p}_4$} into $\tilde{p}_1$ and $\tilde{p}_2$, such that \mbox{$\tilde{p}_{12} = \tilde{p}_1 + \tilde{p}_2 = p_1 + p_2 - \Delta p_{34} = p_{12} - \Delta p_{34}$}. 
		This can be achieved by performing a Lorentz boost in the direction $-\tilde{p}_{12}$ followed by a rescaling with $\sqrt{\tilde{p}_{12}^2/p_{12}^2}$.
	\end{itemize}
We explicitly verified that, in the limit \mbox{$p_{3,T} \to 0 \,(p_{4,T} \to 0)$}, the invariants $\tilde{s}_{13}, \tilde{s}_{23}$ $(\tilde{s}_{14}, \tilde{s}_{24})$ receive corrections proportional to $m^2$, thus screening possibly divergent configurations.
\end{itemize}
In summary, since the $q\bar q$ and $gg$ partonic channels have distinct leading-order momentum flows, we use a dedicated mapping for each channel:
\underline{mapping 1} for $q\bar q$ and \underline{mapping 2} for $gg$, respectively.
The choice of these two mappings resulted from a thorough analysis performed at the one-loop order, where we could validate the quality of the approximation through a direct comparison with the exact results.

\bibliography{biblio}

\end{document}